\documentclass[12pt]{article}
\usepackage[backend=biber,style=apa,uniquename=false]{biblatex}
\usepackage{tabularx}
\usepackage{amsmath}
\usepackage{float}
\usepackage{times}
\usepackage{url}
\usepackage{graphicx}
\usepackage{setspace}
\renewcommand\refname{References and Notes}

\newcounter{lastnote}

\title{Political Fact-Checking Efforts are Constrained by Deficiencies in Coverage, Speed, and Reach}

\author
{Morgan Wack,$^{1\ast}$ Kayla Duskin,$^{2}$ Damian Hodel$^{2}$\\
\\
\normalsize{$^{1}$University of Zurich, Department of Communications and Media Research}\\
\normalsize{$^{2}$University of Washington, Information School}\\
\\
\normalsize{$^\ast$Correspondence email: m.wack@ikmz.uzh.ch}
}

\date{\today}

\begin{document} 
\maketitle 

\begin{abstract}
\singlespacing
\noindent \textit{Fact-checking has been promoted as a key method for combating political misinformation. Comparing the spread of election-related misinformation narratives along with their relevant political fact-checks, this study provides the most comprehensive assessment to date of the real-world limitations faced by political fact-checking efforts. To examine barriers to impact, this study extends recent work from laboratory and experimental settings to the wider online information ecosystem present during the 2022 U.S. midterm elections. From analyses conducted within this context, we find that fact-checks as currently developed and distributed are severely inhibited in election contexts by constraints on their i. coverage, ii. speed, and, iii. reach. Specifically, we provide evidence that fewer than half of all prominent election-related misinformation narratives were fact-checked. Within the subset of fact-checked claims, we find that the median fact-check was released a full four days after the initial appearance of a narrative. Using network analysis to estimate user partisanship and dynamics of information spread, we additionally find evidence that fact-checks make up less than 1.2\% of narrative conversations and that even when shared, fact-checks are nearly always shared within,rather than between, partisan communities. Furthermore, we provide empirical evidence which runs contrary to the assumption that misinformation moderation is politically biased against the political right. In full, through this assessment of the real-world influence of political fact-checking efforts, our findings underscore how limitations in coverage, speed, and reach necessitate further examination of the potential use of fact-checks as the primary method for combating the spread of political misinformation.}
\end{abstract}

\doublespacing

\noindent Widespread use of social media platforms has exacerbated the threat posed to elections by misinformation, with millions of voters subjected to false claims, conspiracies, and unfounded rumors during each election cycle \parencite{coddington2014, guess2018, kennedy2022}. In response to this threat, several interventions have been developed to mitigate the effects of political misinformation \parencite{lazer2018, bovet2019}. Among these, fact-checking, or the ``practice of systematically publishing assessments of the validity of claims" \parencite[p.~251]{walter2020}, has become the most popular strategy for limiting the influence of both apolitical and political misinformation \parencite{clayton2020, zhang2021}. 

The most visible fact-checking is conducted by independent fact-checking organizations (i.e., PolitiFact or Snopes) and by the fact-checking teams associated with mainstream news providers (i.e., Reuters Fact Check, or AP Fact Check). The primacy of these forms of institutional fact-checking to systemic efforts to counter misinformation is reflected in the proliferation of new fact-checking organizations over the past decade \parencite{dukereporterslab2023, graham_increasing_2024}. Despite increased attention and reliance on fact-checking as a critical tool for combating political falsehoods \parencite{marietta2015,ribeiro2021,wasike2023}, less is known about the practical barriers that can reduce overall efficacy in real-world environments \parencite{dias2020, godel_moderating_2021, micallef2022}. As an example, for fact-checks to be effective, they need to be read by their target audiences. Yet, as noted by \textcite{guess_exposure_2020} in their review of engagement with low credibility websites during the 2016 U.S. election, fewer than 3\% of those exposed to an article later determined to contain misinformation also read the corresponding fact-check. Thus far, the challenges posed by practical barriers have yet to be analyzed in full due to limitations with data access \parencite{pasquetto2020,bruns2021,venkatagiri2023}. This gap has required researchers and practitioners to rely on assumptions built on experimental data collected in the absence of observable evidence from active political fact-checking campaigns. 

Critical to our design is the ability to draw on an extensive dataset made up of false, misleading, and uncorroborated election narratives identified on X/Twitter throughout the 2022 U.S. midterm election period \cite{Schafer2024-io}. Unlike prior studies on online misinformation, we start from a baseline set of false and misleading election claims that are not determined by their prior fact-check status. This extension enables us to improve the study's validity while avoiding common challenges related to endogeneity and selection biases. As our conclusions illustrate, this processes enables us to highlight systematic gaps in misinformation research reliant on data from fact-check archives, clarifying an important limitation and potential source of bias present in the literature. 

Most importantly, we not only detail present challenges with selection, but use this new dataset and a secondary process of fact-check identification to test both whether and how three prominent hypothesized barriers reduce their real-world impact. To preview our main contribution, from these campaign-wide analyses we are able to detail how fact-checks are limited by (i.) breadth of coverage - \textit{what} is covered, (ii.) production speeds - \textit{when} it gets covered, and (iii.) messaging reach - \textit{who} sees what is covered. Specifically, we provide evidence that fewer than half of all prominent election-related misinformation narratives were fact-checked. Within the subset of fact-checked claims, we find that the median fact-check was released a full four days after the initial appearance of a narrative. Finally, using clustered partisanship assignment and network visualizations, we find evidence that fact-checks make up less than 1.2\% of narrative conversations and that even when shared, fact-checks are nearly always shared within rather than between partisan communities.

This snapshot of the online information ecosystem challenges previous assumptions and clarifies the limits of institutional fact-checking as it currently functions.

\subsection*{Hypothesized Limitations}

Increased reliance on fact-checking in the political arena has draw attention to their potential limitations. Specifically, researchers have noted limits on the efficacy of fact-checking methods, citing null effects \parencite{garrett2013,jarman2016}, resistance to fact-checks in highly-polarized settings \parencite{flynn2017,bardon2019}, and the concern that fact-checking over-emphasizes the impact of overtly falsifiable forms of misinformation \parencite{amazeen2016,eip2021}. While these criticisms have been used to promote alternatives interventions, fact-checks have largely persisted as the primary method for combating misinformation due to evidence of their effectiveness \parencite{holman2019, porter2019, york2020, bruns2024, porter_correcting_2023}. 

Yet, due in part to persistent challenges inhibiting access to social media data \parencite{pasquetto2020,bruns2021,venkatagiri2023}, assessments of their impact have needed to largely rely on evidence from surveys \parencite{robertson2020,kyriakidou2023,vladika2023,chia2024} and experimental settings \parencite{nyhan2015,graves2016,zhang2021,carnahan2022,koch2023}. As a result, studies often consider fact-checks under ideal conditions which do not account for the real-world factors which may further limit their influence, such as the demand for fact-checks \parencite{graham_increasing_2024}, the timing of fact-checks \parencite{brashier2021}, and exposure to fact-checks \parencite{guess_exposure_2020}. By analyzing the coverage, speed, and reach of fact-checks within the same platform where individuals were exposed to a broader collection of election misinformation from the 2022 U.S. Midterm Elections, our study contributes to the ongoing question of whether the impact of fact-checks ``can be generalized to more naturalistic settings" \parencite[p.~369]{walter2020}.

Finally, in addition to their primary function as a tool for combating misinformation, fact-checking databases composed of extant debunked narratives have been increasingly utilized by researchers to provide validity to the selection of true and false statements. Specifically, where identifying examples of misinformation has presented a hurdle to the conduct of research on misinformation interventions, online fact-check archives have provided researchers with a convenient substitute on which to study the influence of falsehoods. Researchers across disciplines have used fact-checked narratives to compare the spread of true and false rumors online \parencite{vosoughi2018, Friggeri2014-oz, Rosenfeld2020-yt}, as a proxy for misinformation, fake news, or rumors more generally \parencite{allcott2017social, shin2017political, humprecht2019fake}, or to identify potential sources of fake news \parencite{Grinberg2019-wy}. In orienting our work outside of fact-checking datasets we not only improve the study's external validity, but illustrate systematic gaps in commonly used fact-check databases, clarifying an important limitation and potential source of bias present in the literature. 

\subsection*{Present Investigations}

To understand the role of fact-checks in the larger online information ecosystem surrounding the 2022 U.S. midterm elections, we identify election-related rumors as fact-checked or not fact-checked, and subsequently assess how timely and far-reaching online fact-checks were. We draw upon the \textit{ElectionRumors2022} dataset, which provides descriptions and timelines of 135 distinct false, misleading, or unsubstantiated rumor narratives present on X/Twitter surrounding the election \parencite{Schafer2024-io}. To this end, as we do not focus on or attempt to discern intentionality, we rely on a broad conception of ``misinformation'' as an umbrella term which corresponds to any false or misleading information as opposed to more narrowly defined forms of disinformation \parencite{Jack2017-bg}.

This conceptualization allows us to draw on the full dataset, which includes over 1.8 million posts taken from X/Twitter which were judged to have the potential to delegitimize election results or deprive citizens of their vote. As a result, we are able to evaluate the extent of the coverage provided by online fact-checking throughout a U.S. election cycle. By focusing on a highly visible election held in the nation with the largest number of active fact-checking organizations \parencite{Duke2023a}, we view the context of our study as a ``most likely" scenario for the observation of the efficacy of fact-checking efforts.

We divide our findings into three sections for each of the three identified potential barriers to effectiveness. First, we assess what proportion of false, misleading, or unsubstantiated election rumors receive fact-checks, and what factors seem to play a role in selection for fact-checking. We find that less than half of the false, misleading, or unsubstantiated election narratives on X/Twitter during the 2022 election period were fact-checked, and that the fact-checked narratives were not representative of the full set of narratives. Additionally, we show that factors which may be expected to be predictive of whether or not a narrative was fact-checked --- such as narrative virality and influencer involvement --- are less predictive than the type of claim present in the narrative. Next, given that corrections are most effective when presented early, before misconceptions can become entrenched \parencite{vraga2020correction}, we track the speed of fact-check distribution, which is also a central concern of fact-checking researchers and practitioners \parencite{hassan2017,azevedo2018,aslett2021}. We find that fact-checks were, on average, published after 79\% of a narrative's total misinformation posts had already been shared online (a full four days after its appearance on X/Twitter). As a third level of analysis, we assess how fact-checks spread among online audiences, and investigate partisan disparity \parencite{robertson2020,rich2020,graham_increasing_2024}. Using computational and network methods made possible by our unique dataset and collection of fact-checks, we show that even when fact-checks are published and shared on social media, their reach is limited by \textit{who} sees fact-checks due in part to deficient spread across partisan lines. 

\begin{table}
\centering
\caption{Fact-checks by Indicator}
\begin{tabular}{lrrr}
\textit{Narrative Indicator} & \textit{Fact-checked} & \textit{Not Fact-checked} & \textit{Total} \\
\hline
Partisanship: Left-Leaning & 4 (21\%) & 15 (79\%) & 19 \\
Partisanship: Neutral & 0 (0\%) & 3 (100\%) & 3 \\
Partisanship: Right-Leaning & 59 (52\%) & 54 (48\%) & 113 \\
Narrative: Suppression & 6 (19\%) & 26 (81\%) & 32 \\
Narrative: Improbable & 9 (100\%) & 0 (0\%) & 9 \\
Timing: Pre-Election & 28 (37\%) & 47 (63\%) & 75 \\
Timing: Post-Election & 35 (58\%) & 25 (42\%) & 60 \\ 
High Influencer Involvement & 23 (61\%) & 15 (39\%) & 38 \\
High Virality & 14 (54\%) & 12 (46\%) & 26 \\
\hspace{6mm}\textit{Overall Coverage} & 63 (47\%) & 72 (53\%) & 135 \\
\hline
\label{tab:indicators}
\end{tabular}
\begin{minipage}{\linewidth}
\vspace*{-.3cm}
\footnotesize\textmd{Note: \emph{The totals for high influencer involvement (total accounts with 100K followers) and high virality (total posts in the first 24 hours) correspond to narratives with greater than the mean of each variable's total. We rely on this cut-off in the aim of observing how high rates of each variable relate to the comprehensiveness of fact-checking efforts.}}   
\end{minipage}
\end{table}

\section*{Data \& Methods}
To assess the current status of institutional fact-checking within political discourse online, we identify and utilize several sources of relevant data. These include a broad collection of election-related Twitter posts, a high precision dataset of false and misleading election narratives, manually-collected and verified institutional fact-checks relevant to the election, and further enrichment of the Twitter data through estimating the political leaning of users and narratives. The quantitative and qualitative or manual approaches to achieve this are detailed in this section. 

\subsection*{Election Posts}
This work utilizes a large dataset of 446 million posts, which we will refer to as \textit{ElectionTweets2022}. These posts were collected in real time using Twitter's V1.1 API and nine individual collectors between September 5th, 2022 and December 1st, 2022 based on a set of keywords relevant to the U.S. midterm elections (see SI). This dataset was created with the goal of capturing election-related discussion on X/Twitter with high recall. Details on how we use this dataset are in following sections.

\subsection*{False and Misleading Narratives } \label{sec:electionmisinfo}
To conduct our analyses we make use of a dataset detailing false and misleading election narratives active during the 2022 U.S. midterms (\textit{ElectionRumors2022}) \parencite{Schafer2024-io}. This dataset is the result of real-time collection and analyses conducted by a multi-stakeholder group of academic organizations between September 5th and December 1st, 2022. Through observation of social media, instances of uncorroborated, false, or misleading information about the election were identified online, examined, and documented. These instances were qualitatively and quantitatively examined to eliminate duplication, excessively broad, or very low-spread cases. This analysis resulted in the identification of 135 distinct narratives, each centered around a specific claim of false, misleading, or uncorroborated information. 

Additionally, because in this work we analyze the time delay between the first time a narrative is present on X/Twitter and when a fact-check was published, we specifically ensured that the timeline of each narrative was accurate so as not to bias our temporal analyses. To do so, we manually validated that the earliest post from each set of narrative-linked posts identified in \textit{ElectionRumors2022} is indeed related to the targeted narrative.

\subsection*{Fact-check Identification} \label{sec:factcheck_ids}
On top of this dataset of false and misleading election narratives, we conducted a supplementary collection of the published fact-checks related to each rumor. We identified and collected the fact-checks published online that specifically address each misinformation narrative contained in the \textit{ElectionRumors2022} Dataset. 

Given that the majority of Americans rely on Google for news information \parencite{shearer2021news} and place additional trust in the top provided results \parencite{pan2007google}, we replicate factual information-seeking behavior by manually searching Google for fact-checks related to each narrative \parencite{kim2009describing,cotter2022fact}.

While it is possible that some fact-checks are not easily found through querying Google, we consider that discoverability via the search engine to be a good threshold for accessibility of the information. For each narrative, two authors manually and independently created search queries based on the narrative summaries and sampled posts related to that narrative, appending the term "fact check" to a short description of the narrative. Through this process of querying Google, we collected up to six publicly available links that could be fact-checks for each narrative from the first ten results returned by the search engine. Next, we qualitatively evaluated each link based on a detailed set of inclusion criteria. To be included, each link had to meet the following conditions: it must provide an official fact-check (clearly using the term ``fact-check"), directly address the narrative in question (mentioning the location, individual, and the target of the rumored action), be produced after the narrative appeared online (e.g. excluding fact-checks of similar situations from previous elections), and their source had to be determined as unbiased (neither left nor right biased) and of high credibility according to the Media Bias Fact Check (MBFC) ratings \parencite{MBFC2023}. 

Following this procedure, we obtained 164 unique online fact-checks relating to 63 out of the 135 narratives. Once we had identified the online fact-checks, we identified fact-checking posts as any X/Twitter posts present in \textit{ElectionTweets2022} that includes a link to the webpage of the identified fact-check. 

\subsection*{User and Narrative Partisanship Labeling} \label{sec:partisanship}
We estimated the partisanship of each X/Twitter user in our dataset using the \textit{ElectionTweets2022} data. We first identified influencers associated with the political left and political right in election-related discourse. To do so, we used a network-based approach by constructing coengagement networks \parencite{Beers2023-qe} from the reposts in \textit{ElectionTweets2022} to identify left and right leaning influencers. Then, we propagated labels to other users present in the dataset based on who they reposted. For each user present in \textit{ElectionTweets2022}, we totaled the number of reposts of any of the influencers and then labeled the user if 80\% or more of their reposts were of members from a specific partisan cluster. Those that either did not repost any political influencers or who had a more evenly mixed spread of reposts were considered politically neutral. Influencer partisanship labels were then manually validated by the research team to ensure consistency. 

Finally, we identified each narrative as left-leaning or right-leaning. To do so, we reviewed the user-level partisanship participation present in each narrative. Because \textit{ElectionRumors2022} contains some posts that are correcting or countering the related narrative, it is possible that a simple majority of user-participation would not accurately capture the political leaning of the narrative. To navigate this, two subject matter experts reviewed each narrative and designated a left-leaning label when the majority of users \textit{promoting} the narrative were left-leaning. Similarly, any case where the majority of users promoting the narrative were right-leaning received a right-leaning label.
 
\section*{Results: Coverage, Speed, and Reach}

\subsection*{Coverage}
To assess which narratives receive fact-checks, we integrate several variables related to each narrative into our analysis, including partisanship labels, period of occurrence, influencer involvement, virality, and a classification of each variable in a wider election typology, which corresponds to the type of claim identified within each wider narrative. Table \ref{tab:indicators} details the extent of fact-check coverage of unique misinformation narratives (or rumors) by each distinct characteristic throughout the 2022 U.S. Midterm Election period. Most notably, we find that only 47\% (63) of the narratives identified during the election period were ever fact-checked. The remaining 72 narratives were never rebutted. Due in part to the presence of several large narratives, we observe a slightly more inclusive coverage rate of 52\% when considering the percent of all posts stemming from fact-checked narratives. \footnote{mean number of posts per narrative associated with fact-checked narratives was 14,901 compared to 9,637 among narratives that were never fact-checked, while the difference between medians was larger at 9,613 to 1,687, respectively.}

Regarding partisanship, we find that 52\% of right-leaning narratives and only 21\% of left-leaning narratives were fact-checked, with an even larger split observed when considering the total number of posts. This difference in attention supports evidence from other studies which have detailed partisan disparities in fact-check totals related to the 2022 Midterm Elections \parencite{dukereporterslab2023}. However, this gap in coverage between parties attenuates when other variables --namely, the type of claim made within the narrative -- are introduced. Beyond partisanship, we find that narratives arising in the aftermath of the election were fact-checked nearly 20\% more often than pre-election narratives and that narratives with greater than average influencer involvement, as measured by the number of users interacting with a narrative with at least 100,00 followers, were fact-checked only slightly more often than narratives with lower than average numbers of engaged influencers. Similarly, we find only a small difference in coverage when splitting narratives by virality, as defined by the total number of posts produced in the 24 hours following each narrative's first appearance on X/Twitter. Lastly, we find that narratives involving incidents of suppression, or those which focused on issues preventing citizens from voting, were fact-checked less than 20\% of the time, while narratives involving improbable results, or those involving the promotion of statistical anomalies in voting outcomes, were fact-checked in each occasion. 

\begin{figure}[hbt!]
\centering
\includegraphics[width=0.8\linewidth]{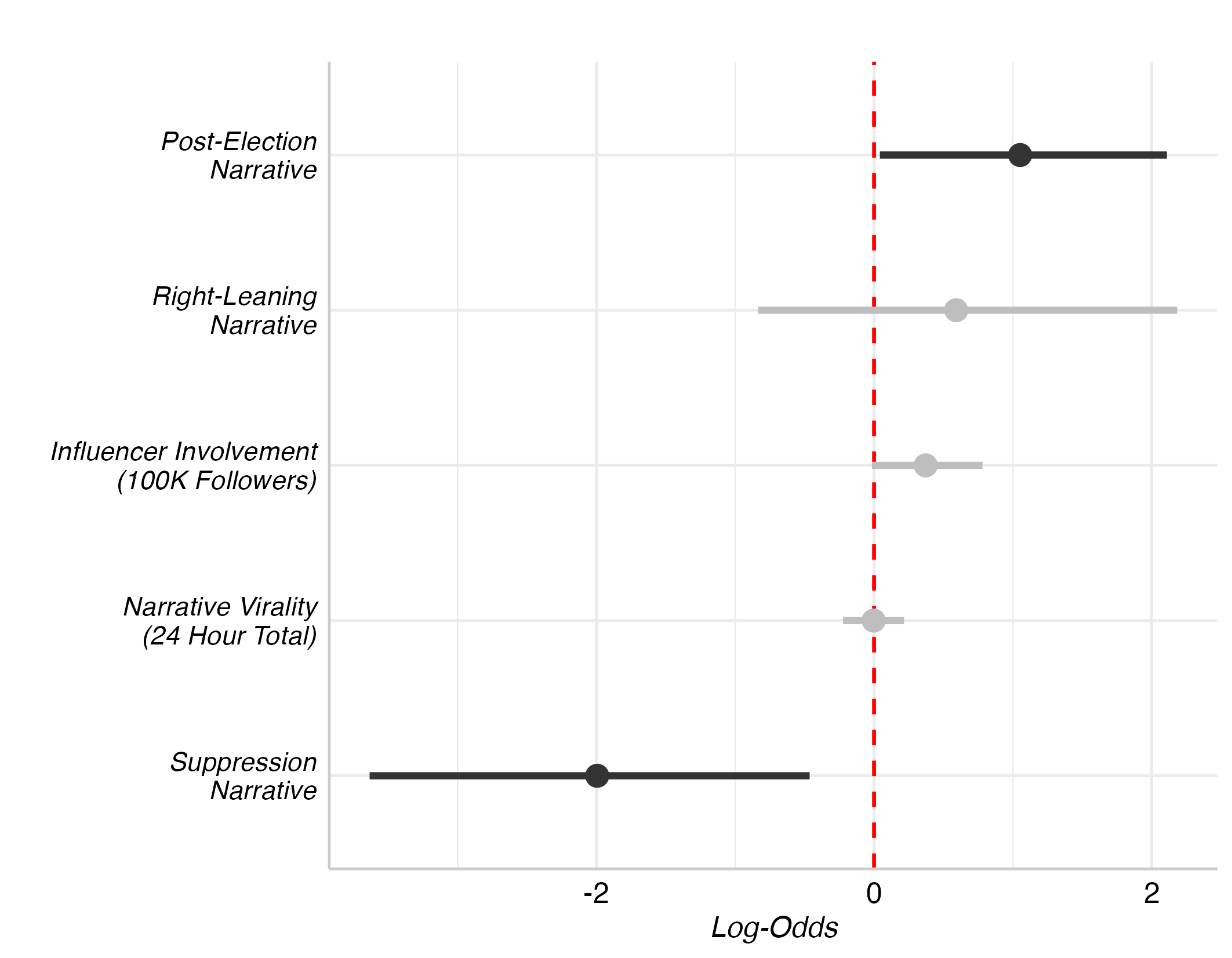}
\caption{\textit{Predicted probability of being fact-checked in the 2022 Midterm Election. Additional classifications are excluded for brevity but included in the SI. Points represent regression coefficients. Lines represent the 95\% confidence interval. The dashed line represents the exact null effect. Black lines are statistically significant at $\alpha = .05$.}}
\label{fig:likelihoodoffactcheck}
\end{figure}

To extend this analysis and probe for gaps in coverage, Figure \ref{fig:likelihoodoffactcheck} details the influence each factor contributed to the likelihood that a specific narrative would or would not be addressed by fact-checkers.\footnote{Additional variables and details regarding the logistic regression output are contained in the Supplementary Materials.}

In accounting for the contributions of each variable, we find that the indicator with the greatest influence on a narrative's propensity to be fact-checked was its time of origin. The greater attention paid to post-election narratives is consistent with the emphasis of fact-checking organizations on monitoring narratives related directly to voting and electoral outcomes. In the other direction, we find a significant disparity in rates of fact-checking by narrative type. Whereas every narrative focused on improbable results was fact-checked, we find comparatively low fact-checking coverage on narratives involving instances of voter suppression. We view these outliers as evidence of an understandable tendency among fact-checking organizations to focus on easily falsifiable information over narratives involving first-person claims. \footnote{Further details on these splits and recommendations for fact-checking organizations to improve claim-based coverage are included in the supplementary materials.} 

When examined as a raw number or as a percentage of narratives, right-leaning rumors appear to be more heavily fact-checked. On the surface this asymmetry may contribute to public perceptions about bias in moderation or ``censorship'' \cite{vogels2020most}. However, consistent with the work of \textcite{jiang2019} on perceived biases in YouTube's moderation, and recent work by \textcite{Mosleh2024-nc} on Twitter moderation, once we control for additional characteristics related to the actual content of the narrative, we find limited evidence that partisanship influenced fact-checking coverage. As a result, studies which rely on fact-checked narratives as the basis for their analyses are likely overstating partisan splits in coverage due to differences in the characteristics of the narratives produced across the simplified left-right political spectrum. Finally, in analyzing the null indicators, we find unexpected results related to narrative virality and influencer engagement. Though we expected the involvement of high follower accounts to receive disproportionately higher coverage due to their reach and prominence in the wider political ecosystem, we find that narratives with higher than average influencer engagement are fact-checked at similar rates to narratives with little influencer engagement. Even more unexpectedly, we find that narrative virality was not a strong predictor of whether a narrative was fact-checked.

In all, in assessing our first barrier we reveal the existence of gaps in coverage which we hope will aid in improving the reach of future fact-checking efforts. Rather than focusing on prominent narratives with high or notable engagement, fact-checking efforts appear to have been highly skewed by feasibility constraints. While this outcome likely comes of no surprise to fact-checkers, who prioritize specific narratives for fact-checking based on their as their human and technological resources, as well as their professional standards \parencite{westlund_what_2024}, the results reiterate the downsides associated with an over-reliance on fact-checking as a central strategy for combating election misinformation. 

\subsection*{Speed}
\label{speed}

When examining the efficacy of debunking initiatives, which includes fact-checking efforts as well alternatives such as credibility labels \parencite{clayton2020}, assessments often treat their application as a binary outcome: they are either present or absent. In the context of elections and other time-sensitive events, the eventual presence of a fact-check is not enough to ensure its influence, it must be timely as well. Speed has become an increasingly challenging barrier due to the accelerating pace of online discourse. To this point, \textcite{pfeffer2023half} estimate that the median post on X/Twitter loses its information value within just 80 minutes. 
Speed has become of even greater concern as large language models (LLMs), such as ChatGPT, are used to amplify online misinformation \parencite{pan_risk_2023}. Although speed has been discussed as a barrier to fact-checking efficacy by researchers \parencite{zhao_insights_2023, beltran_claimhunter_2021, guo_survey_2022}, the impact of slow production speeds has rarely been empirically examined. 

To assess the extent of the timeliness barrier's effect on fact-checking, we examine the response time of fact-checking efforts by tracking the time of the first fact-check to be published on each narrative in our dataset. Using this statistic, we then compare this timing to the progression of each narrative. Critically, we only examine fact-checked narratives. As such, this barrier's influence can be seen as an additional barrier compounding the challenge of coverage.

\begin{figure}[hbt!]
  \centering
  \includegraphics[width=\linewidth]{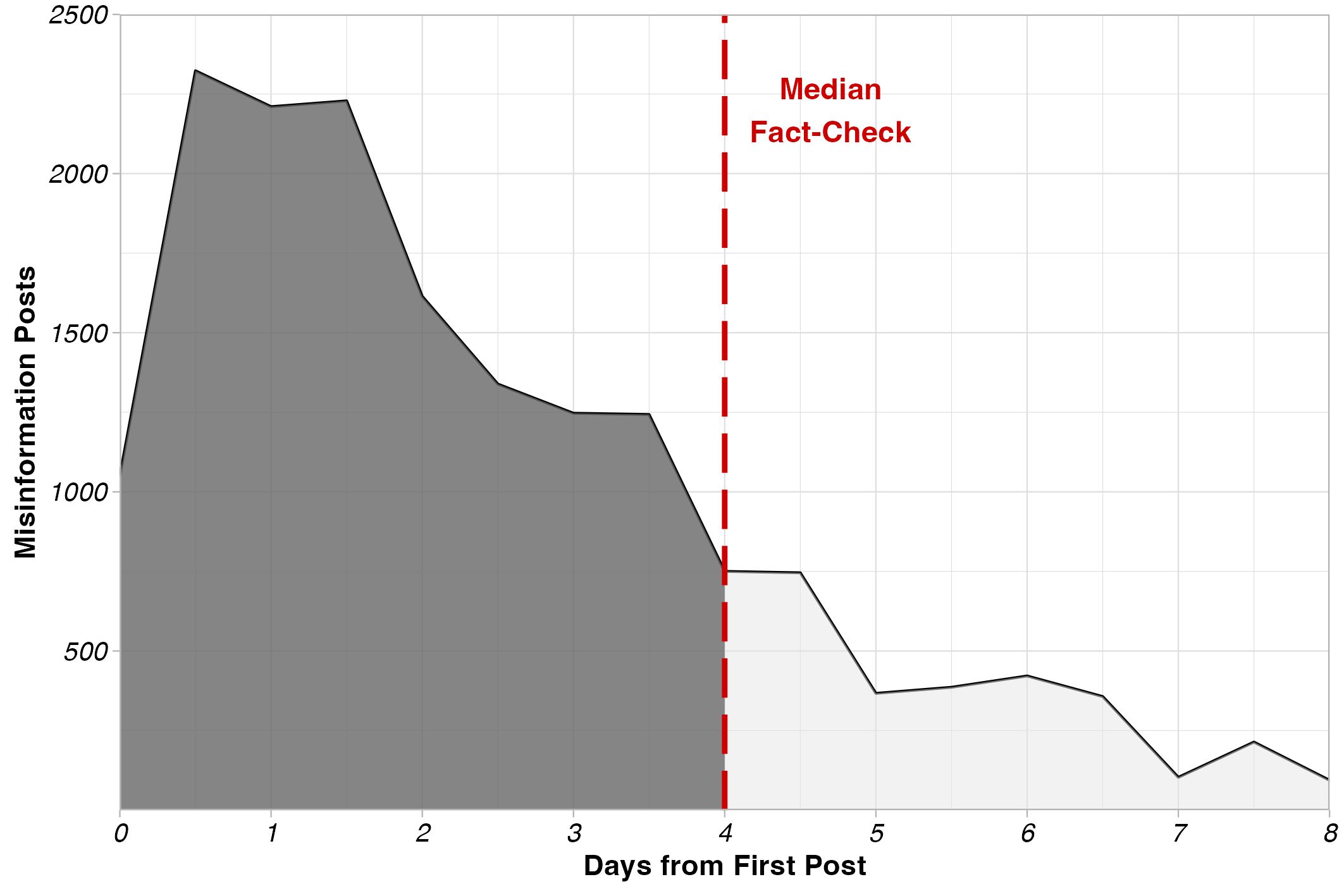}
  \caption{\textit{Misinformation posts published before (black) and after (grey) the modal narrative was fact-checked (grouped by 12 hour totals). Fact-checks most often occur only after the typical rumor has already lost momentum, as indicated by aggregated (median) fact-check response time (vertical line).}}
  \label{fig:ppm_mean_before_after}
\end{figure}%

When aggregated across all 63 fact-checked narratives, we find that the median first fact-check was published a full \textit{four days} after the first appearance of the narrative on X/Twitter, as shown in Figure~\ref{fig:ppm_mean_before_after}. When projected over a longer time frame, we find that for 32\% of the fact-checked narratives, it took over a week for the first fact-check to be produced and published. Of even greater concern, we find that the majority of the discussion related to the modal narrative occurred prior to the publication of the first fact-check. Specifically, we find a median delay of three days between the peak X/Twitter activity for each narrative and the fact-check publication. 
This observation, reflected in the aggregated curve in Figure~\ref{fig:ppm_mean_before_after}, suggests that the modal fact-check (red) most often occurred after the rumor was already losing momentum. As a result, at the time of the median first fact-check's publication a significant proportion (79\%) of posts had already been posted suggesting that the drop-off was unrelated to the publication of the fact-check itself.\footnote{see the \textit{Coverage \& Speed} section in the SI.} 

\begin{figure*}
\centering
\includegraphics[width=\linewidth]{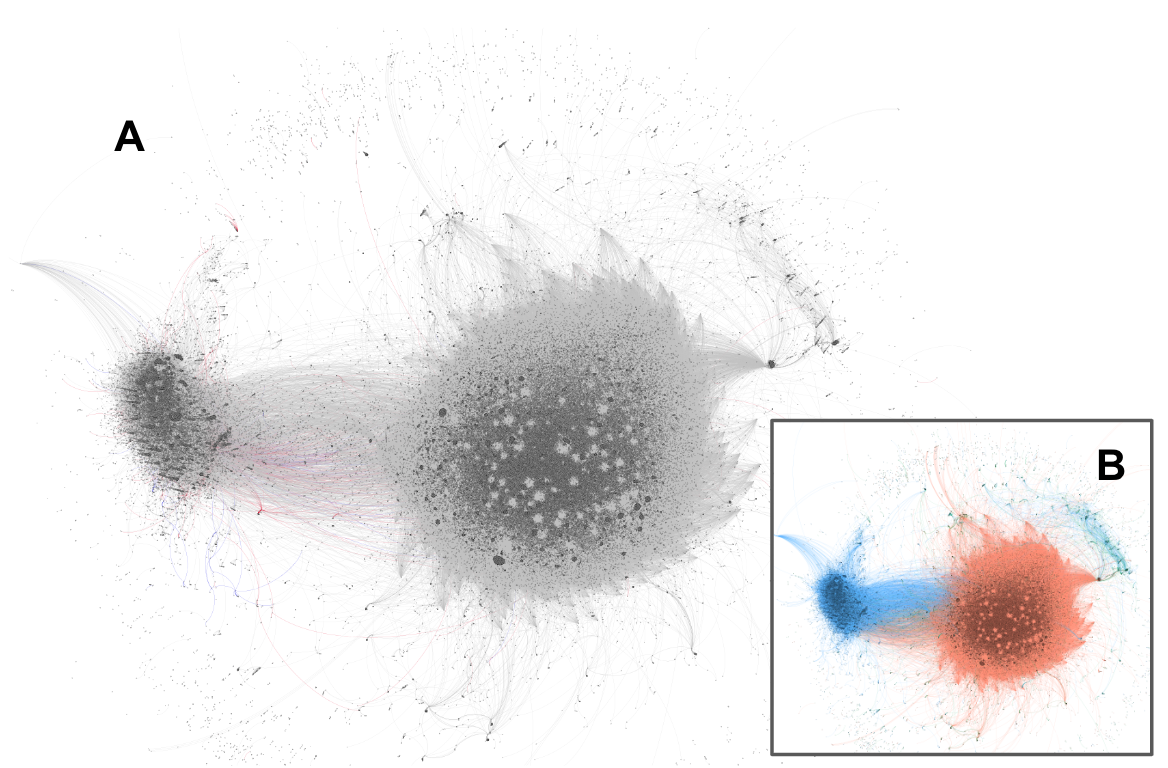}
\caption{\textit{Network visualization of online discussion of false and misleading narratives and their corresponding fact-checks. Nodes represent X/Twitter users while edges represent reposts between users. In Figure A, nodes are all colored gray while edges are colored based on the content of the repost. Red edges are fact-checks of right-leaning narratives while blue edges are fact-checks of left-leaning narratives. Gray edges are all other narrative-related reposts that are not fact-checks. In Figure B, nodes are colored by the partisanship of the user (light blue is left-leaning, light red is right-leaning, green is neutral) and edges are colored according to the partisanship of the source (reposting) user. The left cluster is primarily left-leaning users while the larger cluster on the right is right-leaning. Visualization conducted using the ForceAtlas2 graph layout algorithm on Gephi network visualization software.}}
\label{fig:network}
\end{figure*}

\subsection*{Reach}

For fact-checks to influence perceptions of information veracity, they not only need to be published and produced quickly, but also widely circulated \parencite{burel2021}. Recent evidence has suggested the production of fact-checks has overshadowed attention by organizations and researchers on this fundamental constraint, ensuring that ``the main factors that can facilitate or hinder the spread of fact-checks on the internet remain research gaps and need in-depth investigation" \parencite[p. 1482]{li2023}. To address this additional data deficit, the final barrier we examine is fact-check reach, or the extent to which published fact-checks actually extend into the election information ecosystem online. To assess the reach of fact-checks within the platform where rumors were spreading, we identified all election-related posts that include a link to a relevant fact-check, which, as noted, already subsets the full set of misinformation to include only covered narratives.

Even within this targeted subset of fact-checked narratives, posts including fact-checks are dwarfed by the total volume of posts related to the targeted misinformation narratives. In full, we identified 12,510 fact-check posts among the 1,045,373 posts that are not fact-checks. Put differently, posts containing or linking to fact-checks made up 1.2\% of the conversation related to these narratives. In further interrogating the links between these posts, we found that users participating in false and misleading narratives were highly unlikely to interact with posts containing fact-checks. Specifically, only 4.15\% of left-leaning users in our dataset shared a fact-check while right-leaning and politically neutral users were even less likely to share a fact-check, at 0.30\% and 0.38\%, respectively. Figure \ref{fig:network} visualizes these disparities between the false, misleading or unsubstantiated narratives and the corresponding corrective information in the wider network of X/Twitter posts about election rumors. 

In addition to the relative volume of fact-checking posts, we also assess whether fact-checks spread among the same communities that initially shared the false or misleading narratives. We first categorize each narrative as left or right leaning (see \textit{Materials \& Methods}). The partisan makeup of users posting about left and right wing narratives and their respective fact-checks are shown in Table~\ref{tab:party_participation}. General discussion on the narrative topics align in terms of narrative partisanship and participant partisanship with 97\% of posts about left-leaning narratives posted by left-leaning users and 74\% of posts about right-leaning originated from right-leaning users. Fact-checks, on the other hand, do not follow the same trend. Left-leaning users post 85\% of the fact-checks focused on left-leaning narratives and 80\% of the fact-checks related to right-leaning narratives. While left-leaning accounts appear to fact-check themselves and the opposing party with high frequency, the same is not true of right-leaning accounts. Right-leaning users were responsible for 10\% of fact-checking posts focused on right-leaning narratives and 9\% of fact-checking posts on left-leaning narratives. 

\begin{figure*}[hbt!]
\centering
\includegraphics[width=\textwidth]{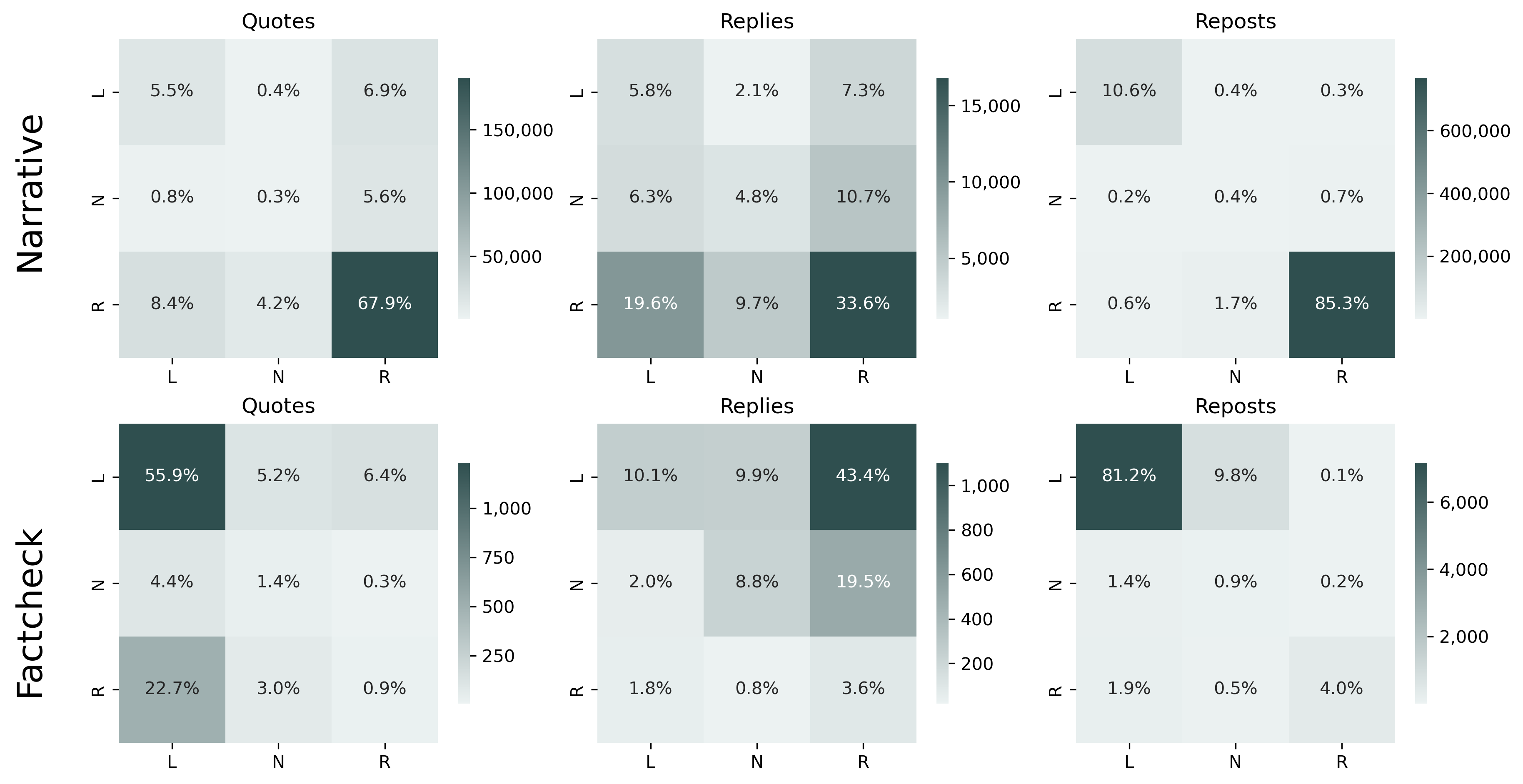}
\caption{\textit{Mechanisms of spread between users with different party leanings for posts discussing the false and misleading narratives and posts sharing fact-check links. Y-axis denotes the party (L - left-leaning, N - neutral/apolitical, R - right-leaning) of the posting user (e.g. the one who quotes, reposts, or replies), X-axis denotes the party of the original poster.}}
\label{fig:interaction_heatmap}
\end{figure*}

Finally, we observe \textit{how} the different types of information spread within these communities. We consider the three main ways that users can publicly interact with one another on X/Twitter -- reposting, quote posting, and replying. Figure \ref{fig:interaction_heatmap} shows that narratives primarily spread from right-leaning user to right-leaning user through all three forms of interaction, though replies are somewhat split with 34\% of replies coming from right-leaning users replying to right-leaning users and 20\% of replies coming from right-leaning users replying to left-leaning users. Consistent with the aforementioned findings, we see that fact-checks were most often shared from left-leaning user to left-leaning user through quote posts and reposts. Replies that contain fact-checks, however, were most likely to come from a left-leaning user replying to a right-leaning user (43\%) or a neutral user replying to a right-leaning user (20\%). While 23\% of quote posts that contain fact-check links were right-leaning users quoting left-leaning users, there is hardly any spread from right-leaning users to other right-leaning users across any modality of post. 

\begin{table}
    \caption{\textit{Partisan participation by narrative alongside fact-checks of left-leaning and right-leaning narratives}}
    \centering
    \begin{tabular}{l r r|r r}  
         &  \multicolumn{2}{c|}{\textit{Narrative Posts}}&  \multicolumn{2}{c}{\textit{Fact-check Posts}}\\ 
         User Alignment &  Left Narr.&  Right Narr&  Left Narr& Right Narr\\ \hline 
        Left-Leaning & 97.25\% & 15.63\% & 84.86\% & 79.64\% \\
        Neutral & 1.46\% & 9.96\% & 6.15\% & 10.04\% \\
        Right-Leaning & 1.29\% & 74.41\% & 8.99\% & 10.32\% \\
         Total& 32,228 & 292,248 & 1,268 & 6,935 \\ \hline
    \end{tabular}
    \label{tab:party_participation}
\end{table}


\section*{Discussion}
Recent work on fact-checking has illustrated their potential for mitigating the potential harms associated with with misinformation. However, as we have detailed, practical barriers can severely reduce the likelihood that readers will encounter relevant fact-checks in the real-world. The results of these collected analyses have implications for future efforts to combat misinformation and protect trust in democratic elections. The conclusions emphasize the need for greater investment in understanding the environmental factors present in ecologically valid settings which may limit the influence of fact-checking efforts beyond experimental settings.

The results of our study indicate that even when fact-checking \textit{can} minimize the influence of misinformation in principle, there is no guarantee that it \textit{will} in practice, as the real-world barriers present in political settings serve to substantially reduce the scope of fact-checking efforts. While each barrier is connected, they also pose individual challenges which require innovative solutions to ensure that fact-checking is a viable strategy for addressing misinformation. In all, fewer than half of identified misinformation narratives were fact-checked at all. Of the half which were fact-checked, the average fact-check was published a full four days after the first related misinformation post. Lastly, even when fact-checks were shared on social media, fewer than 2\% of users who engage with false and misleading narratives also shared a fact-check associated with that debunked narrative. 

Given these identified limits to the overall coverage of fact-checking efforts, the current findings are important for several reasons. First, through in situ analysis of online rumors and fact-checks, this work illustrates the need for further research on misinformation which draw from realistic data samples rather than unrepresentative subsets. Rather than relying on fact-checks as the starting place, or denominator, the use of broader online data reveals the disparities that persist in coverage, such as the under-representation of suppression narratives, while also illustrating why perceptions of partisan bias persist. Second, in incorporating longitudinal data linked to narrative-based posts, we illustrate the how widely misinformation circulates prior to the publication of associated fact-checks, which are often not produced until the modal narrative is already far beyond the point of peak virality. Finally, even in circumstances in which each of these prior issues is addressed, we detail how partisan disparities in the sharing of fact-checks in the U.S. political context minimize their potential audience and likely impact overall efficacy. 

The results detailed here provide insight into specific challenges that we hope practitioners will be able to utilize. Specifically, we generate evidence that highlights new areas for fact-checking and misinformation research efforts while emphasizing the need to extend existing work which acknowledges practical barriers to impact. Our finding regarding coverage first reiterate the need for research which analyzes misinformation beyond solely the narratives which are eventually fact-checked. Secondly, we encourage coordination among fact-checking organizations and the introduction of a more diverse set of public entities in the fact-checking process. This could help extend what is covered during election periods, since, as detailed further in the supplementary information, we find that fact-checks are often duplicated across organizations due in part to competition for viewership among fact-checking organizations \parencite{amazeen2016}, potentially reducing the breadth of coverage. These insights should be of particular interest to the journalists involved in the production and publication of fact-checks, the users who share the fact-checks on social platforms, and the platform operators who can influence the dissemination of fact-checks, e.g. through platform design. In addition, given the need for speed in addressing false rumors, our work provides further support for efforts aimed at automating certain portions of the corrective process \parencite{beltran_claimhunter_2021}. The potential benefits of automated detection and amplification include the potential to minimize both time lost to production and dissemination efforts. Finally, the insulation of fact-checks to a relatively small group of users speaks to the challenges of breaking through echo-chambers present on social media \parencite{Nikolov2021-rh}. This highlights the importance of research in increasing the demand for fact-checking \parencite{graham_increasing_2024} and the production of correction crafted to appeal to diverse audiences \parencite{saveski2022engaging}.

In addition to these insights, our work can also be used to both justify and interpret the potential of alternative methods for combating misinformation. Specific to misinformation on X/Twitter, the introduction of in-situ fact-checking efforts, including the community notes program, show promise as an efficient and effective misinformation mitigation strategy \parencite{Borwankar2022-gx, Saeed2022-ht}.  While the Community Notes program was being reconfigured as a successor to ``Birdwatch" during November, 2022, our findings support in principle the potential for fact-checks generated within a platform to aid in minimizing the deleterious effects of the barriers identified in this work. Namely, depending on implementation, community-based fact-checking have the potential to extend the reach of corrections beyond the targets of organizations, improve the speed of correction applications, and bypass partisan-specific distributions where notes are attached directly to the post rather than relying on users for their circulation. While the program shows potential \parencite{Drolsbach2024-to}, it is also necessary to note that crowd-sourced methods have been found to be most helpful when linking to trustworthy sources \parencite{Prollochs2022-cs}, a role that traditional fact-checking organizations have the capability to fill. In all, while our work lends support to the extension of both community notes and similar efforts in principle, it remains to be seen how the program will evolve to understand its eventual impact on the elimination of practical corrective barriers.

With these improvements in mind, we also must acknowledge the limitations and potential challenges to the generalizability of our outcomes. First, our study is specific to the U.S. electoral context. While the attention afforded to political events such as elections likely increases fact-checking activity, it also applies additional pressure on fact-checking organizations to maximize their impact in a short period of time. Whether fact-checkers are more likely to have improved coverage, speed, or reach in alternative settings remains an open question in need of further research. Moreover, we suggest caution in the interpretation of our results to elections outside the U.S., given that it is the country with the most official fact-checking organizations, and is home to a specific brand of electoral politics which cannot be generalized to geographies with distinct political and social histories. Second, we center this work on one specific social media platform. While X/Twitter has historically been the center of conversation on U.S. political news \parencite{stier2020}, it is not the only way that people encounter instances of misinformation or their fact-checks. As online attention is increasingly fractured across social media platforms \parencite{auxier2021social}, new studies will be needed to assess if the findings of this study hold. As greater attention is shifted toward the practical barriers limiting the influence of fact-checking, we hope our work is extended to address a more diverse set of topical and geographic cases. Finally, we acknowledge that while our analyses detail the severity of extant challenges to fact-checking efforts, they do not address the full set of corrective interventions available to online platforms, including most notably textual corrections which do not link to an external fact-check as well as crowd-sourced alternatives.

\newpage
\singlespacing
\printbibliography

\noindent \textbf{Acknowledgements:} 
The authors thank Joseph Schafer for lending expertise and help with the \textit{ElectionRumors2022} data.\\

\noindent \textbf{Funding:} This material is based upon work supported by:

The University of Washington's Center for an Informed Public 

The John S. and James L. Knight Foundation (G-2019-58788)

The William and Flora Hewlett Foundation (2022-00952-GRA)

The Election Trust Initiative

The National Science Foundation (grant no. 2120496).
\\
\noindent Additionally, K.D. would like to acknowledge support by the National Science Foundation Graduate Research Fellowship under Grant No DGE-2140004. Any opinions, findings, conclusions, or recommendations expressed in this material are those of the authors and do not necessarily reflect the views of the National Science Foundation or other funders.\\

\noindent \textbf{Author Contributions} \\

\noindent M.W, K.D., and D.H designed researched, performed research, analyzed data, and wrote the paper. \\



\noindent \textbf{Data and materials availability:} \\

\noindent All data and code used for the development of the analyses included in the main text and SI will be made publicly available for the purposes of reproduction and extension. Specifically, in addition to the publication of the accompanying dataset used for the identification of 2022 U.S. election narratives \parencite{Schafer2024-io}, we will publish data on the fact-checks associated with each narrative constructed by the authors. 

\clearpage





\renewcommand\refname{References and Notes}





\subsection*{Data Description}
\label{SI:electiontweets}
\subsubsection*{ElectionTweets2022}
Posts on X/Twitter were collected in real time using Twitter’s V1.1 Academic streaming API. Keywords were engineered to capture a very broad set of posts, emphasizing recall over precision. Keywords were initiated in July 2022 and were periodically updated as relevant topics emerged. The final set of keywords is listed here: 

\begin{table}[h!]
    \centering
    \begin{tabularx}{\textwidth}{|X|}\hline 
        `BMD', `BMDs', `EVM', 'EVMs', `HandMarkedPaperBallots', `USPS', `absentee', `adjudication', `arizona, `audit', `ballot', `ballots', `bmd', `bmds', `chain of custody', `chicago', `cisa', `cochise', `code', `codes', `color revolution', `decertification',`decertify',`desantis',`detroit',`dominion',`drop box',`drop boxes',`dropbox',`dropboxes', `election',`election2022',`electioneering',`electionfraud', `elections', `elections2022', `electors', `electors','epoll', `es\&s', `forensic',`fortalice', `fraud', `fraudulent', `fulton', `halderman', `hand count',`hand-count',`hand-counted',`hand-marked', `handcount', `handcounted', `handcounts', `handmarked', `imagecast', `imagecastx', `integrity', `intimidation', `lancaster',`machine', `machines', `mail', `mail in', `maricopa', `michigan', `midterm', `midterms',  `midterms2022', `mule', `nomachines', `noncitizen', `onenightcount', `overvote', `paperballots',  `pennsylvania', `philadelphia', `pima', `pinal', `poll', `pollbook', `pollbooks', `pollbooks', `polling', `polls', `pollwatcher', `pollwatchers', `pollworker', `post office', `postal', `postoffice', `precinct', `racine', `raffensperger', `recount', `results', `rigged', `rigged voterfraud', `risk-limiting audit',  `rolls', `smartmatic', `subversion', `suppression', `tabulator', `tabulators', `tallies', `tamper',  `tampered', `tampering', `touchscreen', `touchscreens', `undervote', `vote', `votebymail', `voted', `voter', `voterfraud', `voters', `votersuppression', `votes', `votesuppression', `voting', `vulnerabilities', `vulnerability', `yuma' 
        \\ \hline
    \end{tabularx}
    \label{tab:keywords}
\end{table}

\newpage
\subsection*{Coverage Analysis}

To assess the probability of having been fact-checked in the 2022 U.S. Midterm Elections we rely on a binomial logistic regression based on the following formula:


\begin{equation}
\begin{aligned}
\text{Pr}(\text{Fact-check} = 1) &= \beta_0 + \beta_1 \times \text{Partisanship} \\
&+ \beta_2 \times \log(\text{Virality}) \\
&+ \beta_3 \times \log(\text{Influencers}) \\
&+ \beta_4 \times \text{Classification} \\
&+ \beta_5 \times \text{Timing}
\end{aligned}
\end{equation}

\noindent \textit{where} the output represents the binary outcome of whether a fact-check was produced or the specified narrative remained un-fact-checked (0/1), $\beta_0$ is the intercept, $\beta_1$ is a categorical variable with three levels denoting variations partisan leanings among the narratives (right-leaning, neutral, and left-leaning), $\beta_2$ represents a logged count of the total posts related to the narrative produced in the first ten hours (virality rate), $\beta_3$ represents a logged count of the number of users with at least 100,000 followers who promoted the narrative (influencer count), $\beta_4$ denotes the classification of misinformation coded for the narrative in question, while $\beta_5$ is a binary variable denoting whether the first post in the narrative was prior to (0) or on/following (1) election day (November 8th, 2022). The odds ratios corresponding to this equation and partially displayed in Figure 1 in the main paper are reflected in the table below. Subsequent subsections detail the predicted probability of each variable over an extended range. For data type, we expand on the definitions and explain the coding process. 

\begin{table}[!htbp] \centering 
  \caption{} 
\begin{tabular}{@{\extracolsep{5pt}}lc} 
\\[-1.8ex]\hline 
\hline \\[-1.8ex] 
 & \multicolumn{1}{c}{\textit{Dependent Variable:}} \\ 
\cline{2-2} 
\\[-1ex] & \\ 
 & Likelihood of Fact-Check \\ 
\hline \\[-1.8ex] 
 Partisanship: Neutral & $-$16.036$^{***}$ \\ 
  & (1.239) \\
 Partisanship: Right-Leaning & 0.592 \\ 
  & (0.920) \\ 
 Virality (logged) & $-$0.003 \\ 
  & (0.119) \\ 
 Influencer Engagement (logged) & 0.372 \\ 
  & (0.254) \\ 
 Classification: Corruption & $-$0.616 \\ 
  & (0.959) \\ 
 Classification: Dilution & 0.032 \\ 
  & (0.716) \\ 
 Classification: Elimination & $-$1.019 \\ 
  & (0.860) \\ 
 Classification: Improbable & 16.484$^{***}$ \\ 
  & (0.832) \\ 
 Classification: Manipulation & $-$0.005 \\ 
  & (1.095) \\ 
 Classification: Obstruction & $-$0.636 \\ 
  & (0.954) \\ 
 Classification: Suppression & $-$1.994$^{**}$ \\ 
  & (0.894) \\ 
 Classification: Suspicion & $-$0.234 \\ 
  & (0.983) \\ 
 Period: Post-Election Day & 1.052$^{*}$ \\ 
  & (0.546) \\  
 Constant & $-$1.377 \\ 
  & (1.196) \\ 
\hline \\[-1.8ex] 
Observations & 135 \\ 
Log Likelihood & $-$70.127 \\ 
Akaike Inf. Crit. & 168.255 \\ 
\hline 
\hline \\[-1.8ex] 
\textit{Note:}  & \multicolumn{1}{r}{$^{*}$p$<$0.1; $^{**}$p$<$0.05; $^{***}$p$<$0.01} \\ 
\end{tabular} 
\end{table} 

\newpage
\subsection*{Influence of Influencers (100K+) on Fact-Check Probability}

\begin{figure}[h!]
\centering
\includegraphics[width=.8\linewidth]{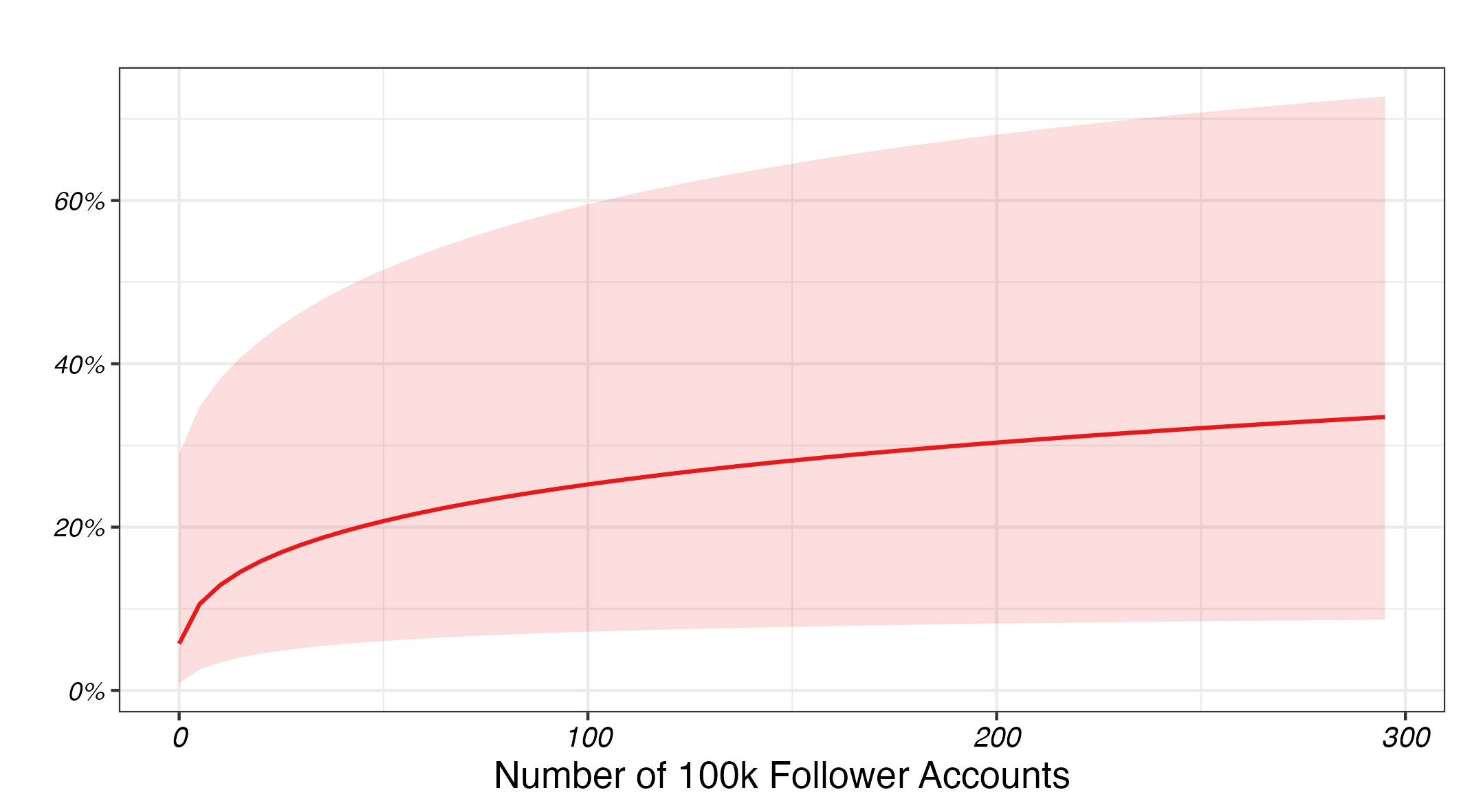}
\caption{\textit{The plot illustrates the expected influence of influence involvement (total users with 100K+ followers) on Fact-Check probability.}}
\label{fig:follower_plot}
\end{figure}

\noindent Figure 1 shows the marginal effects of the influencer engagement variable (number of 100K follower accounts per narrative) on the likelihood that a fact-check is produced. All other variables are held at their mean/modal values. The slight upward bend of the trend line details the positive, but insignificant relationship between influencer engagement and fact-check probability. 

\newpage
\subsection*{Influence of Virality Rate (<24 hrs) on Fact-Check Probability (log)}

\begin{figure}[h!]
\centering
\includegraphics[width=.8\linewidth]{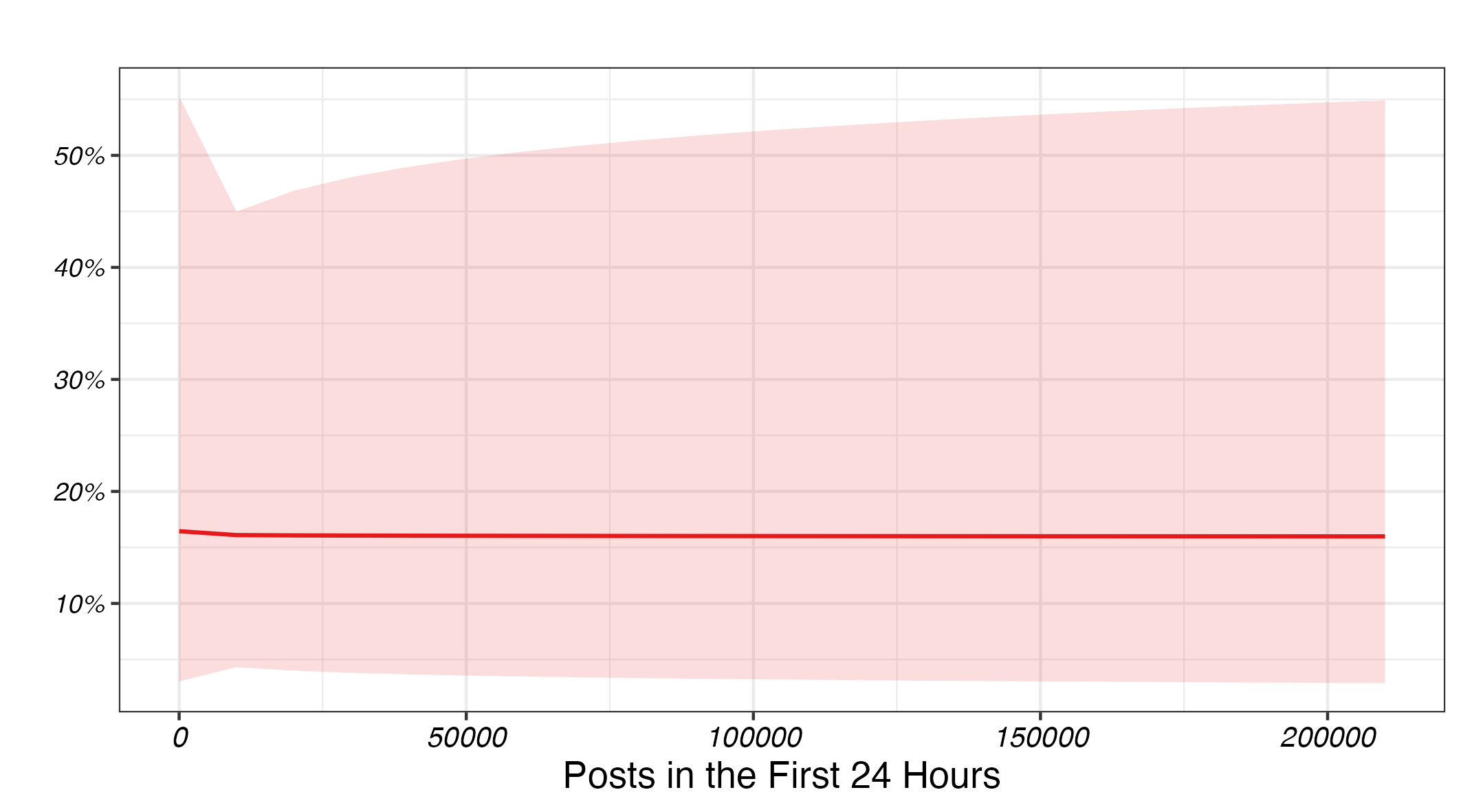}
\caption{\textit{The plot illustrates the expected influence of narrative virality (<24 hrs) on Fact-Check probability.}}
\label{fig:fc_citations}
\end{figure}

\noindent Figure 2 shows the marginal effects for the narrative virality variable (number of posts per narrative in the first day/24 hours) on the likelihood that a fact-check is produced. All other variables are held at their mean/modal values. The horizontal trend line reflects the insignificant relationship between narrative virality and fact-check probability observed in the data. 

\newpage
\subsection*{Influence of Partisanship of Narrative on Fact-Check Probability}

\begin{figure}[h!]
\centering
\includegraphics[width=.8\linewidth]{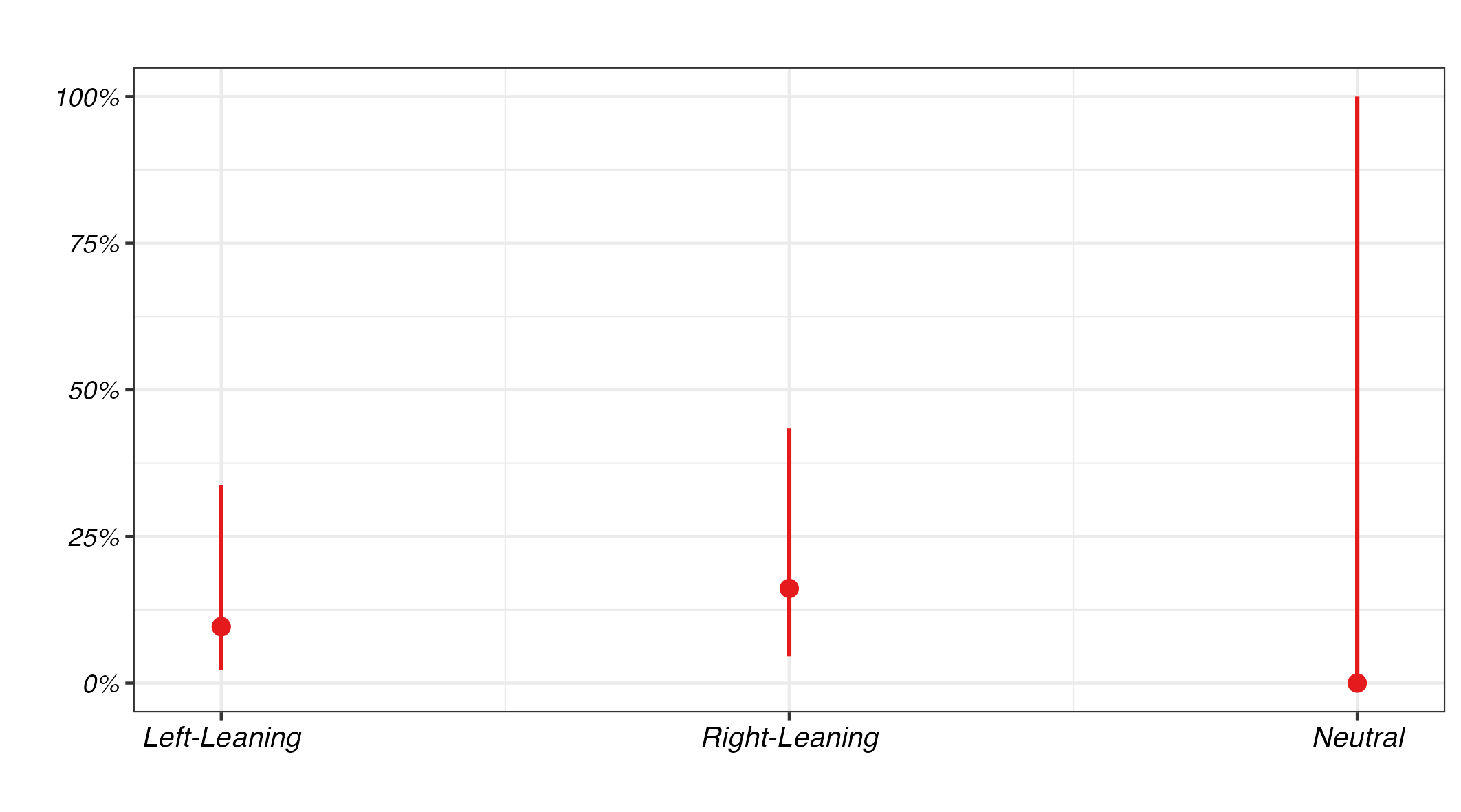}
\caption{\textit{The plot illustrates the expected influence of partisan alignment on fact-check probability. Numeric variables are held at their mean while factor variables are assigned their modal value from the dataset.}}
\label{fig:partisan_plot}
\end{figure}

\noindent Figure 3 details the marginal effects associated with narrative partisanship (see \textit{Materials \& Methods} for more details on partisan classification) on the likelihood that a fact-check is produced. All other variables are held at their mean/modal values. As observed in the logistic regression output, right-leaning narratives are slightly more likely to be fact-checked, though the difference between right- and left-leaning is minimal. Both exhibit low rates of coverage. As none of the three neutral narratives were fact-checked, the corresponding marginal output is difficult to interpret substantively. 
\newpage

\subsection*{Influence of Narrative Timing on Fact-Check Probability}

\begin{figure}[h!]
\centering
\includegraphics[width=\linewidth]{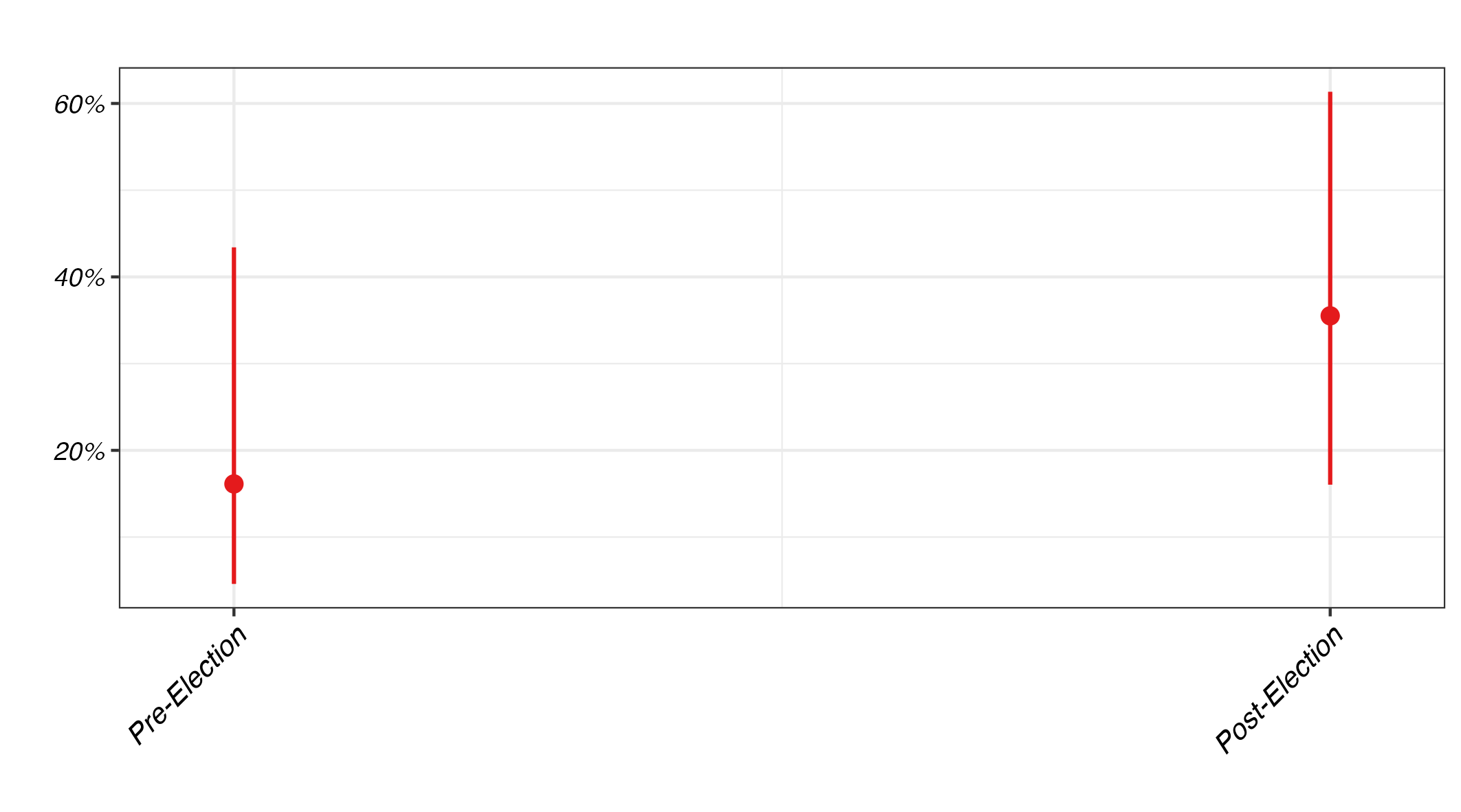}
\caption{\textit{The plot illustrates the expected influence of the timing of a narrative's appearance on fact-check probability. Numeric variables are held at their mean while factor variables are assigned their modal value from the dataset.}}
\label{fig:temporal_plot}
\end{figure}

\noindent Figure 4 illustrates the marginal effects associated with the timing of each narrative on the likelihood that a fact-check is produced. To correspond with the efforts of fact-checking organizations, which have been shown to coordinate resources around election day and its aftermath, we use the end of the day prior to election day - 11-7-2022 - as the cut-off between periods. All other variables are held at their mean/modal values. Consistent with the attention and resources expended by fact-checking organizations, post-election narratives appear slightly more likely to be fact-checked than pre-election narratives. As noted in the paper containing further information on the underlying dataset, this differential in coverage prior to the election could be used to inform future fact-checking efforts in the aim of improving their overall efficiency given persistent financial and staff-based constraints. 

\newpage
\subsection*{Types of Election Misinformation}
To determine the types of misinformation included in the analyses, the dataset was coded based on Mike Caulfield's work categorizing common tropes related to election misinformation. Each narrative in the \textit{ElectionMisinfo2022} Dataset was coded based on the nine categories detailed here. 

\noindent The first set of codes relates to the broader category of "Changed Results". This includes:

\begin{itemize}
\item \textbf{Suppression} - A voter who would have voted is stopped from voting through some legal or extralegal means (in our set, incidents should be in the extra-legal category)
\item \textbf{Manipulation} - A voter is confused, coerced, or intimidated into casting their vote in a way that does not reflect their intent.
\item \textbf{Elimination} - A vote that the voter has completed is discarded, slow-walked, lost, or unfairly adjudicated in a way that the vote is never tabulated.
\item \textbf{Alteration} - A vote is altered, misread, changed electronically, or adjudicated in a way that is contrary to what the voter intended. This can include methods to confuse the voter into voting for someone else.
\item \textbf{Dilution} - An invalid or fraudulent set of votes is allowed/injected into the system, thus diluting the value of valid votes. Examples can be fraudulent ballots added in, non-citizens being allowed to vote, dead voters on the rolls. 
\end{itemize}

\noindent The second category of common electoral misinformation is focused on "Improbable Results/Timing". This section includes:

\begin{itemize}
\item \textbf{Improbable Results} - No specific claim is made about an action or process. Rather it is claimed that some result or statistical pattern is evidence of fraud. Includes improbable timing.
\item \textbf{Suspicious Timing} - The speed of the process, or the pattern of the results as they come in is evidence of fraud.
\end{itemize}

\noindent The final category is focused on "Untrustable Results". This section includes:

\begin{itemize}
\item \textbf{Improper/Obstructed Oversight} - It is claimed the election is being conducted in such a way that the results are not fact-checkable, or that proper oversight of the election is being blocked.
\item \textbf{Corruption} - It is claimed that the people in charge of various aspects of the election are corrupt, connected to nefarious interests, or have too much impact on the results. 
\end{itemize}
\newpage

\noindent The probability of a narrative coded as any one of these codes is displayed in the following figure: 

\begin{figure}[h!]
\centering
\includegraphics[width=\linewidth]{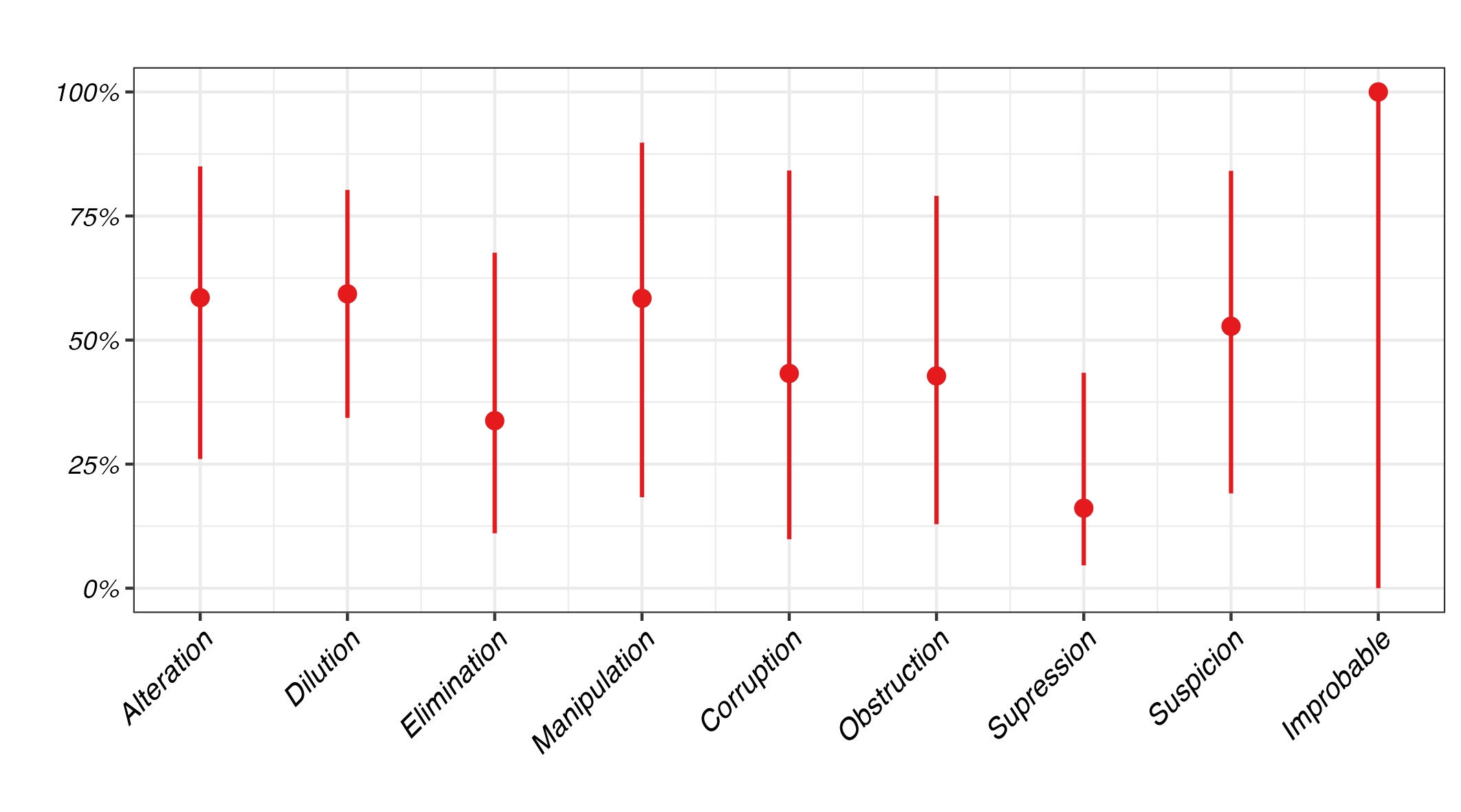}
\caption{\textit{Classification of Narratives}}
\label{fig:classification}
\end{figure}

\noindent In viewing the results, the primary takeaways can best be viewed in the contrast between "Improbable" narratives and "Suppression" narratives. All ten narratives coded as "Improbable" were fact-checked in our dataset, while only six of thirty-two coded as "Suppression" were fact-checked. The splits in the marginal effects outputs displayed in Figure 5 illustrate disparities that result in differential attention and coverage of specific types of misinformation. Suppression narratives, which are often less specific and at times difficult to confirm, are not well-suited for the current organization of fact-checking efforts. On the other hand, narratives involving improbable results both draw attention and often necessitate only easily obtained evidence to debunk. Future work should dive further into these splits to aid in improving the efficacy of the management of fact-checking resources.

\newpage
\subsection*{Coverage and Speed}
\label{sec:appendix:speed}

\begin{figure}[h!]
\centering
\includegraphics[width=\linewidth]{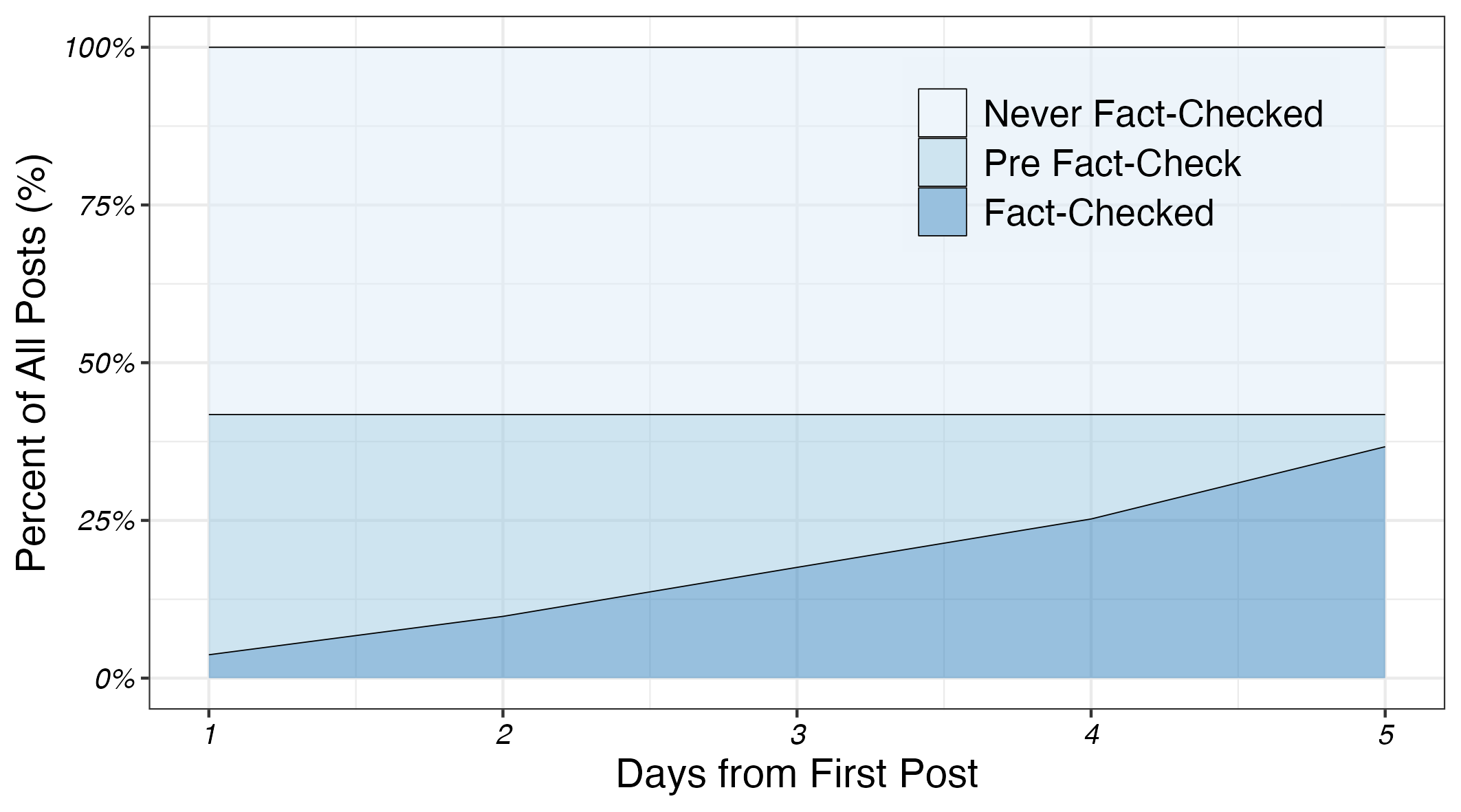}
\label{fig:Percent_of_Posts}
\end{figure}

Figure~\ref{fig:Percent_of_Posts} illustrates our findings and the estimated losses associated from the first two barriers: Coverage and Speed.

\noindent Figure~\ref{fig:Percent_of_Posts} visualizes the extent to which structural barriers minimize coverage from fact-checking over time. In illustrating the cumulative coverage of fact-checking, the above image visualizes the proportion of misinformation which was either never covered, or covered at a later point, in comparison to misinformation which was introduced following the production of an associated fact-check over the first five days since the publication of the first misinformation post for each narrative.


\newpage
\begin{figure}[h!]
\centering
\includegraphics[width=\linewidth]{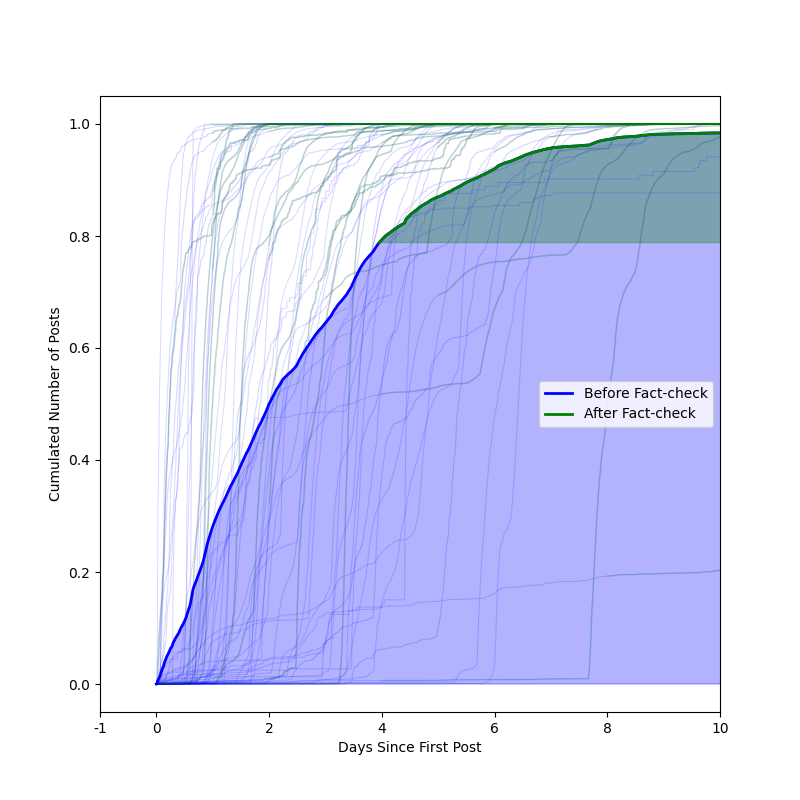}
\caption{\textit{Cumulative posts, normalized between 1 and 0, before (in blue) and after (in green) the publication of the first fact-check. The bold line represents the mean across all fact-checked narratives.}}
\label{fig:median_cumsum_normalized}
\end{figure}

\noindent Figure~\ref{fig:median_cumsum_normalized} shows the cumulative number of posts aggregated across all fact-checked narratives, both before and after the first fact-check for each narrative. Notably, a substantial proportion of posts had already been published by the time of the first fact-check publication.

\newpage
\begin{figure}[h!]
\centering
\includegraphics[width=\linewidth]{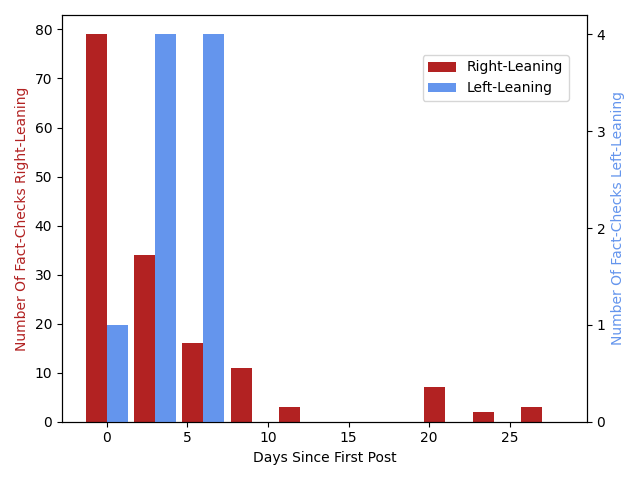}
\caption{\textit{Number of fact-checks categorized into 3-day intervals, divided by partisanship.}}
\label{fig:responsetimedist_fcs}
\end{figure}

\begin{figure}[h!]
\centering
\includegraphics[width=\linewidth]{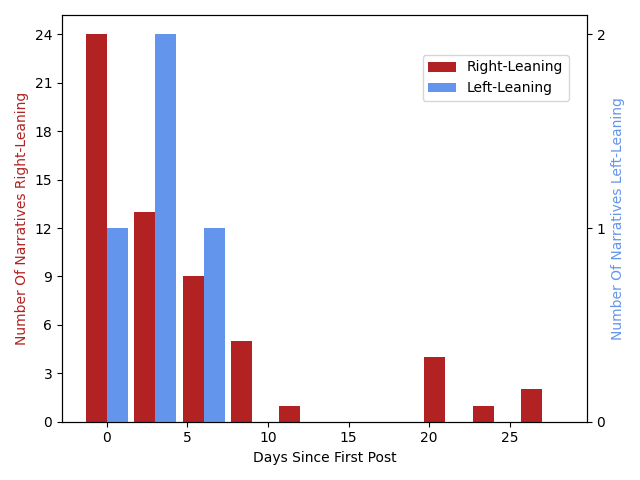}
\caption{\textit{Number of narratives categorized into 3-day intervals, divided by partisanship.}}
\label{fig:responsetimedist_inc}
\end{figure}
\newpage 

\noindent Figure~\ref{fig:responsetimedist_fcs} and Figure~\ref{fig:responsetimedist_inc} illustrate the distribution of the publication times of fact-checks, categorized by partisanship, in terms of both the number of fact-checks and narratives, respectively. 

\subsection*{Fact-checks Per Narrative}

\begin{figure}[H]
\centering
\includegraphics[width=\linewidth]{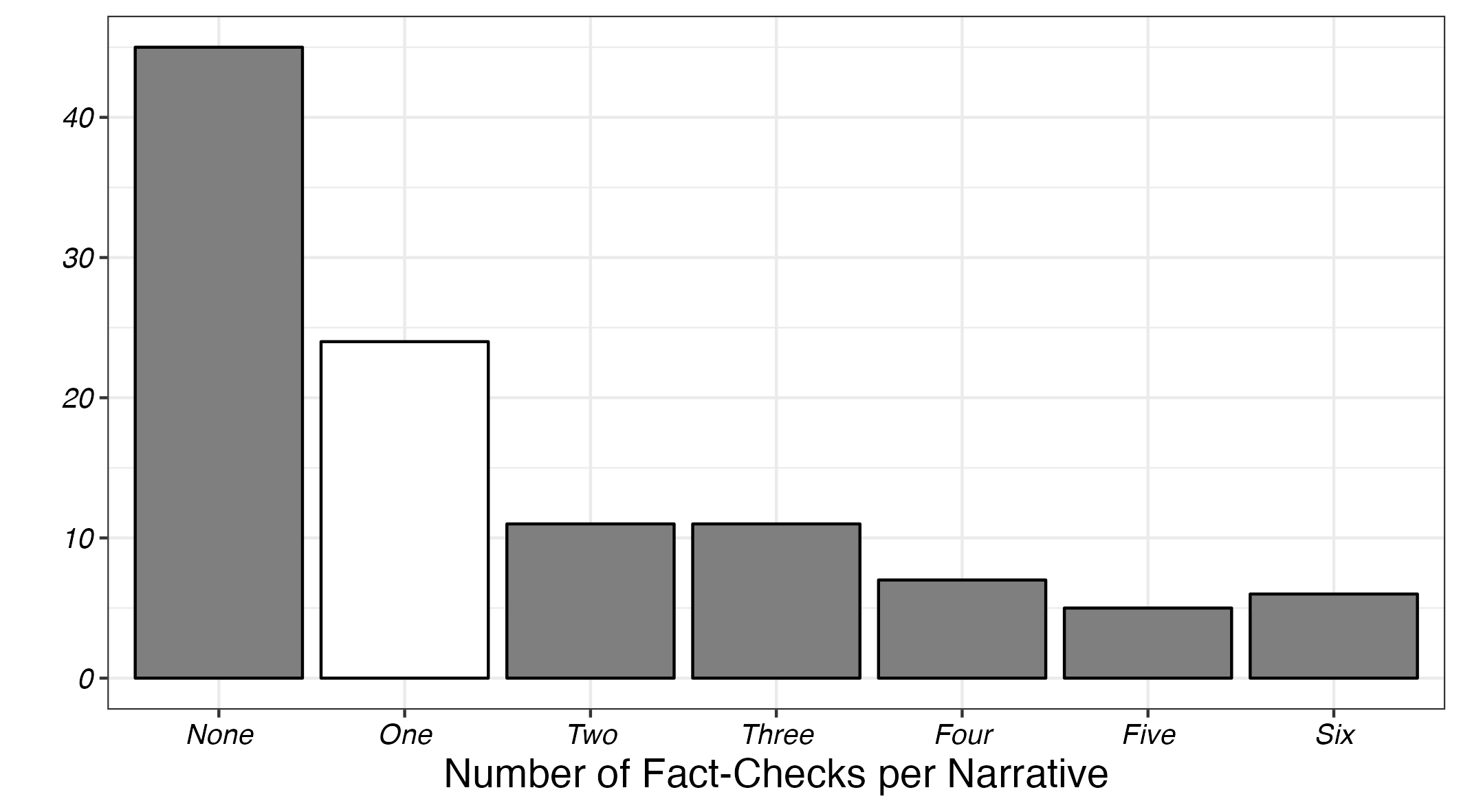}
\caption{\textit{Total Fact-checks for Each Narrative in the Dataset}}
\label{fig:FCs_Narrative}
\end{figure} 

\noindent Duplicated efforts were common in the dataset. As seen in Figure~\ref{fig:FCs_Narrative} only 22\% of the narratives were fact-checked by a single organization. While the ideal number of fact-checks may be greater than one if each subsequent fact-check is addressing a separate portion of the claim or reaching a different audience, this does not appear to be a common occurrence. Rather, it appears that the most prominent narratives and the narratives which were easiest to debunk attracted attention from several organizations at the same time. Future work should improve coordination among organizations to minimize duplication and maximize resources. 

\subsection*{Alternative Variable Cut-offs}

\noindent To ensure that the selected cut-offs for each variable are not driving results we detail the results across separate cut-offs for the virality and influencer variables. 

\begin{table}[H] \centering 
  \caption{} 
\begin{tabular}{@{\extracolsep{5pt}}lcc} 
\\[-1.8ex]\hline 
 & \multicolumn{2}{c}{\textit{Dependent Variable:}} \\ 
\cline{2-3} 
 & Likelihood of Fact-Check &  \\ 
\\[-1.8ex] & (10 Hrs) & (24 Hrs)\\ 
\hline \\[-1.8ex] 
 Partisanship: Neutral & $-$15.994$^{***}$ & $-$16.036$^{***}$ \\ 
  & (1.313) & (1.239) \\ 
 Partisanship: Right-Leaning & 0.587 & 0.592 \\ 
  & (0.967) & (0.920) \\ 
 10 Hour Virality (logged) & $-$0.105 &  \\ 
  & (0.117) &  \\ 
 24 Hour Virality (logged) &  & $-$0.003 \\ 
  &  & (0.119) \\ 
 Influencer Engagement (logged) & 0.428$^{*}$ & 0.372 \\ 
  & (0.226) & (0.254) \\ 
 Classification: Corruption & $-$0.792 & $-$0.616 \\ 
  & (0.949) & (0.959) \\ 
 Classification: Dilution & 0.077 & 0.032 \\ 
  & (0.710) & (0.716) \\ 
 Classification: Elimination & $-$1.041 & $-$1.019 \\ 
  & (0.860) & (0.860) \\ 
 Classification: Improbable & 16.371$^{***}$ & 16.484$^{***}$ \\ 
  & (0.784) & (0.832) \\ 
 Classification: Manipulation & 0.023 & $-$0.005 \\ 
  & (1.107) & (1.095) \\ 
 Classification: Obstruction & $-$0.707 & $-$0.636 \\ 
  & (0.970) & (0.954) \\ 
 Classification: Suppression & $-$2.001$^{**}$ & $-$1.994$^{**}$ \\ 
  & (0.897) & (0.894) \\ 
 Classification: Suspicion & $-$0.338 & $-$0.234 \\ 
  & (0.993) & (0.983) \\ 
 Period: Post-Election Day & 1.308$^{**}$ & 1.052$^{*}$ \\ 
  & (0.562) & (0.546) \\ 
 Constant & $-$1.213 & $-$1.377 \\ 
  & (1.232) & (1.196) \\  
\hline \\[-1.8ex] 
Observations & 135 & 135 \\ 
Log Likelihood & $-$69.574 & $-$70.127 \\ 
Akaike Inf. Crit. & 167.149 & 168.255 \\ 
\hline 
\hline \\[-1.8ex] 
\textit{Note:}  & \multicolumn{2}{r}{$^{*}$p$<$0.1; $^{**}$p$<$0.05; $^{***}$p$<$0.01} \\ 
\end{tabular} 
\end{table} 

\noindent The results of the main logit regression are reproduced with virality levels assigned at different hour totals (10 hours and 24 hours). Results persist.

\newpage

\begin{table}[H] \centering 
  \caption{} 
\begin{tabular}{@{\extracolsep{5pt}}lcc} 
\\[-1.8ex]\hline 
\hline \\[-1.8ex] 
 & \multicolumn{2}{c}{\textit{Dependent Variable:}} \\ 
\cline{2-3} 
 & Likelihood of Fact-Check &  \\ 
\\[-1.8ex] & (100K) & (250K) \\ 
\hline \\[-1.8ex] 
 Partisanship: Neutral & $-$16.036$^{***}$ & $-$16.288$^{***}$ \\ 
  & (1.239) & (1.129) \\ 
 Partisanship: Right-Leaning & 0.592 & 0.452 \\ 
  & (0.920) & (0.848) \\ 
 Virality (logged) & $-$0.003 & 0.054 \\ 
  & (0.119) & (0.112) \\ 
 100K Influencer Involvement (logged) & 0.372 &  \\ 
  & (0.254) &  \\ 
 250K Influencer Involvement (logged) &  & 0.245 \\ 
  &  & (0.261) \\ 
 Classification: Corruption & $-$0.616 & $-$0.401 \\ 
  & (0.959) & (0.933) \\ 
 Classification: Dilution & 0.032 & 0.037 \\ 
  & (0.716) & (0.713) \\ 
 Classification: Elimination & $-$1.019 & $-$0.999 \\ 
  & (0.860) & (0.861) \\ 
 Classification: Improbable & 16.484$^{***}$ & 16.771$^{***}$ \\ 
  & (0.832) & (0.805) \\ 
 Classification: Manipulation & $-$0.005 & $-$0.008 \\ 
  & (1.095) & (1.104) \\ 
 Classification: Obstruction & $-$0.636 & $-$0.527 \\ 
  & (0.954) & (0.962) \\ 
 Classification: Suppression & $-$1.994$^{**}$ & $-$1.963$^{**}$ \\ 
  & (0.894) & (0.871) \\ 
 Classification: Suspicion & $-$0.234 & $-$0.086 \\ 
  & (0.983) & (0.965) \\ 
 Period: Post-Election Day & 1.052$^{*}$ & 0.974$^{*}$ \\ 
  & (0.546) & (0.535) \\ 
 Constant & $-$1.377 & $-$1.183 \\ 
  & (1.196) & (1.116) \\ 
\hline \\[-1.8ex] 
Observations & 135 & 135 \\ 
Log Likelihood & $-$70.127 & $-$71.267 \\ 
Akaike Inf. Crit. & 168.255 & 170.534 \\ 
\hline 
\hline \\[-1.8ex] 
\textit{Note:}  & \multicolumn{2}{r}{$^{*}$p$<$0.1; $^{**}$p$<$0.05; $^{***}$p$<$0.01} \\ 
\end{tabular} 
\end{table} 

\noindent The results of the main logit regression are reproduced with influencer account levels assigned at different levels of followers (100K and 250K). Results persist.

\subsection*{OLS Coverage Analysis}

Finally, we recreate the main coverage analysis with an OLS to ensure that the results of the logit persist across alternative specifications. The equation for the model is defined as follows:


\begin{equation}
\begin{aligned}
\text{Fact-check}_i = &= \beta_0 + \beta_1 \times \text{Partisanship} \\
&+ \beta_2 \times \log(\text{Virality}) \\
&+ \beta_3 \times \log(\text{Influencers}) \\
&+ \beta_4 \times \text{Classification} \\
&+ \beta_5 \times \text{Timing}  + \epsilon_i
\end{aligned}
\end{equation}

\noindent In this model, the dependent variable represents the binary outcome of whether a narrative was fact-checked (1) or remained un-fact-checked (0). Here, $\beta_0$ is the intercept, $\beta_1$ is a categorical variable representing variations in partisan leanings among the narratives with three levels (right-leaning, neutral, and left-leaning), $\beta_2$ denotes the log-transformed count of the total posts related to the narrative produced in the first ten hours (indicative of virality rate), $\beta_3$ captures the log-transformed count of the number of users with at least 100,000 followers who promoted the narrative (reflecting influencer involvement), $\beta_4$ is a categorical variable that classifies the narrative according to misinformation types, and $\beta_5$ is a binary variable indicating whether the first post in the narrative occurred prior to (0) or on/following (1) election day. The coefficient estimates from the OLS regression, which reflect the expected change in probability of a fact-check for a one-unit increase in the predictor variables, are presented in the Figure and Table below. Here we are able to visualize the "improbable" classification unlike in the logit output. 

\newpage

\begin{figure}[H]
\centering
\includegraphics[width=.8\linewidth]{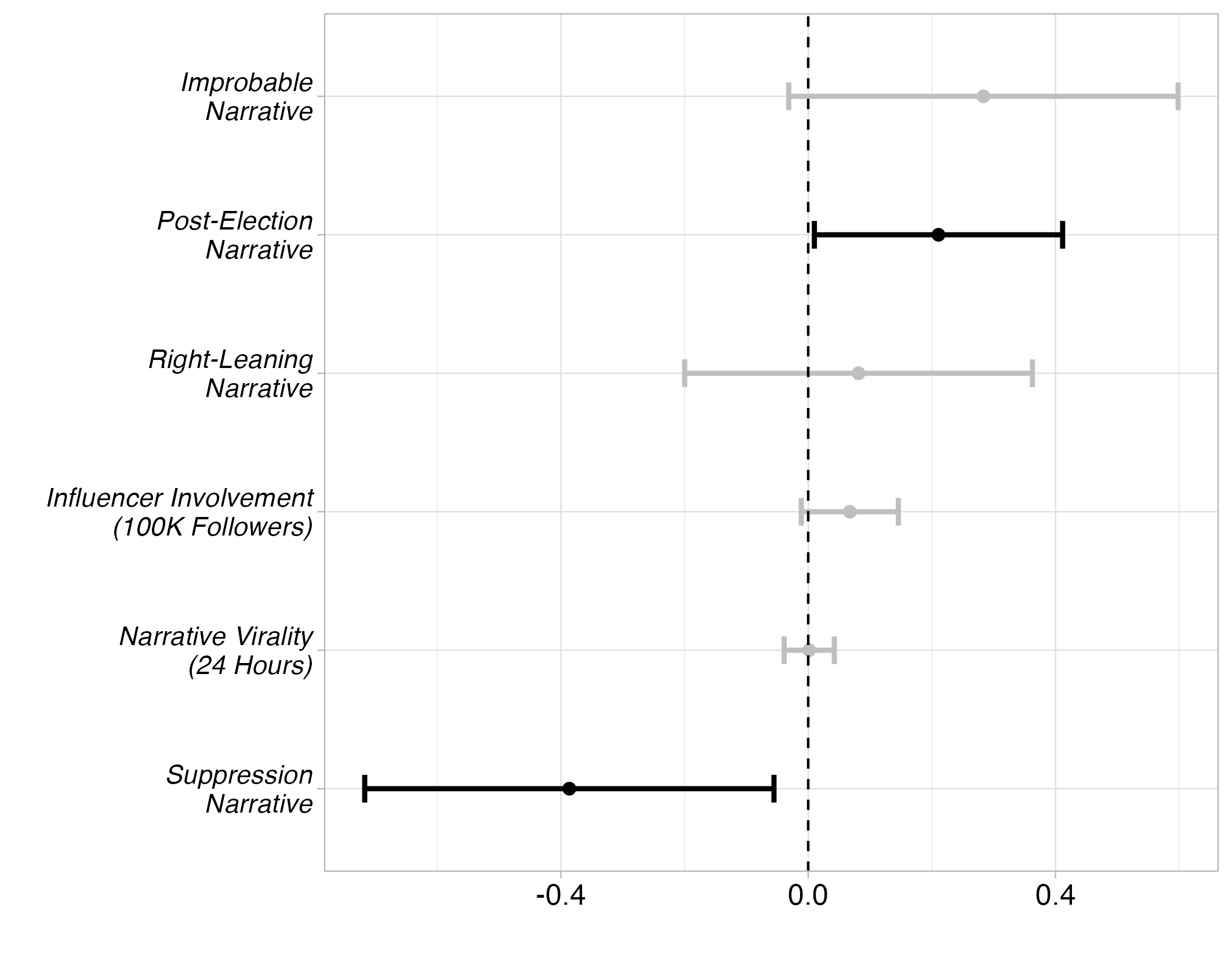}
\caption{OLS Recreation of Main Logit Analysis}
\label{fig:ols_plot}
\end{figure}

\begin{table}[H] \centering 
  \caption{} 
\begin{tabular}{@{\extracolsep{5pt}}lc} 
\\[-1.8ex]\hline 
\hline \\[-1.8ex] 
 & \multicolumn{1}{c}{\textit{Dependent Variable:}} \\ 
\cline{2-2} 
 & Change in Probability of Fact-Check \\ 
\hline \\[-1.8ex] 
Partisanship: Neutral & $-$0.208 \\ 
  & (0.159) \\ 
 Partisanship: Right-Leaning & 0.092 \\ 
  & (0.142) \\ 
 Virality (logged) & 0.013 \\ 
  & (0.019) \\ 
 Influencer Engagement (logged) & 0.055 \\ 
  & (0.039) \\ 
 Classification: Corruption & $-$0.094 \\ 
  & (0.198) \\ 
 Classification: Dilution & $-$0.002 \\ 
  & (0.160) \\ 
 Classification: Elimination & $-$0.224 \\ 
  & (0.177) \\ 
 Classification: Improbable & 0.301$^{**}$ \\ 
  & (0.150) \\ 
 Classification: Manipulation & $-$0.033 \\ 
  & (0.244) \\ 
 Classification: Obstruction & $-$0.118 \\ 
  & (0.207) \\ 
 Classification: Suppression & $-$0.382$^{**}$ \\ 
  & (0.166) \\ 
 Classification: Suspicion & $-$0.041 \\ 
  & (0.203) \\ 
 Period: Post-Election Day & 0.190$^{**}$ \\ 
  & (0.093) \\ 
 Constant & 0.231 \\ 
  & (0.216) \\ 
\hline \\[-1.8ex] 
Observations & 135 \\ 
Adjusted R$^{2}$ & 0.209 \\ 
Residual Std. Error & 0.445 (df = 121) \\ 
F Statistic & 3.719$^{***}$ (df = 13; 121) \\ 
\hline 
\hline \\[-1.8ex] 
\textit{Note:}  & \multicolumn{1}{r}{$^{*}$p$<$0.1; $^{**}$p$<$0.05; $^{***}$p$<$0.01} \\ 
\end{tabular} 
\end{table} 


\end{document}